\documentclass[conference]{IEEEtran} 
\IEEEoverridecommandlockouts

\usepackage{cite}
\usepackage{amsmath,amssymb,amsfonts}
\usepackage{graphicx}
\usepackage{textcomp}
\usepackage{xcolor}
\def\BibTeX{{\rm B\kern-.05em{\sc i\kern-.025em b}\kern-.08em
    T\kern-.1667em\lower.7ex\hbox{E}\kern-.125emX}}
\usepackage{amsmath}
\usepackage{amssymb}
\usepackage{graphicx}
\usepackage{booktabs}
\usepackage{algorithm}
\usepackage{algpseudocode}
\usepackage{caption}
\usepackage{cite}
\usepackage{comment} 
\usepackage{authorindex}
\usepackage{subcaption}

\begin{document}

\title{Component Centric Placement Using Deep Reinforcement Learning}

\author{\IEEEauthorblockA{Kart Leong Lim \\
 Institute of Microelectronics (IME), \\
 \textit{Agency for Science, Technology and Research (A{*}STAR),}\\
 2 Fusionopolis Way, Innovis \#08-02, Singapore 138634, Republic of
Singapore \\
 limkl@a-star.edu.sg}}

\maketitle

\begin{abstract}
Automated placement of components on printed circuit boards (PCBs) is a critical stage in placement layout design. While reinforcement learning (RL) has been successfully applied to system-on-chip IP block placement and chiplet arrangement in complex packages, PCB component placement presents unique challenges due to several factors: variation in component sizes, single- and double-sided boards, wirelength constraints, board constraints, and non-overlapping placement requirements. In this work, we adopt a component-centric layout for automating PCB component placement using RL: first, the main component is fixed at the center, while passive components are placed in proximity to the pins of the main component. Free space around the main component is discretized, drastically reducing the search space while still covering all feasible placement; second, we leverage prior knowledge that each passive's position has to be near to its corresponding voltage source. This allows us to design the reward function which avoids wasted exploration of infeasible or irrelevant search space. Using the component centric layout, we implemented different methods including Deep Q-Network, Actor-Critic algorithm and Simulated Annealing. Evaluation on over nine real-world PCBs of varying complexity shows that our best proposed method approaches near human-like placements in terms of wirelength and feasibility.
\end{abstract}

\section{Introduction}

PCBs and chiplets share fundamental similarities, with the primary challenge being the spatial arrangement of functional blocks to minimize interconnect length, reduce crosstalk, and meet thermal constraints. Placement strategies can be broadly categorized as analytical \cite{cheng2018replace,lin2019dreamplace}, partitioning \cite{sait1999vlsi,855354}, simulated annealing \cite{chen2006modern,kirkpatrick1983optimization,chen2005modern} and more recently, reinforcement learning based methods \cite{cheng2021joint,mirhoseini2021graph}. Applying RLs to PCB placement is challenging due to several factors: algorithms must support both single- and double-sided boards with different constraints; extend from single- to multi-chip layouts with added connectivity; define reward functions that balance wirelength, congestion, and feasibility; handle diverse component sizes within the state and action space; and ensure non-overlapping placement for manufacturable designs. In this work we target discrete PCB placement. A set of discrete physical locations arranged around a main component, allowing each component to occupy a valid position in the layout. To efficiently explore this placement space, we combine reinforcement learning with a component-centric layout strategy: the main component is fixed at the center, while passive components are positioned near power pins to minimize wirelength and prevent overlaps. This exploration driven AI placement generates novel layouts which not only satisfy design constraints but also incorporate component centric intent into the placement process.

\subsection{Discrete Action RLs}

Deep Q-Networks (DQN) \cite{mnih2013playing,sutton1998reinforcement,schaul2015prioritized} provide deterministic and efficient learning by directly maximizing Q-values, making them well-suited for discrete action spaces and typically stable during training. Their limitations, however, are notable: DQN is restricted to discrete actions, cannot naturally extend to continuous action spaces, and may be less flexible in environments where stochastic exploration or policy diversity is beneficial. Actor-Critic (AC) methods \cite{mnih2016asynchronous,NIPS1999_6449f44a} address some of these shortcomings by combining value-based and policy-based learning. The critic estimates state or action values to stabilize updates, while the actor explicitly optimizes the policy, enabling effective learning in both discrete and continuous action spaces. Nonetheless, AC methods also face challenges, as rewards are learned indirectly through the critic's estimates, which can make policy optimization slower and less direct; in addition, in discrete action spaces, the actor's probability distribution over all actions may introduce unnecessary stochasticity. In contrast to both DQN and AC, Decision Transformers (DT) \cite{chen2021decision} approach reinforcement learning from a sequence modeling perspective. By conditioning on past states, actions, and the return-to-go, DTs can leverage expert trajectories effectively and achieve stable performance without the instability of trial-and-error optimization. They also allow flexible conditioning on target returns, offering a degree of controllability absent in traditional methods. However, DTs are fundamentally limited by their reliance on pre-collected trajectories: they cannot learn from random or exploratory placements outside the dataset, lack the ability to interact with the environment to improve beyond demonstration quality, and require large, diverse datasets to generalize effectively across placement scenarios.

\section{Proposed Method}

Component centric layout is a design strategy embraced by some PCB designs, in which passive components are deliberately organized around a central or main component - such as microcontrollers, power circuits, and connectors. Passives are strategically positioned close to power pins to minimize wirelength and maintain non-overlaps between passives, to reflect an intentional design philosophy. We apply the component-centric design strategy to RL in two key ways:

\begin{itemize}
\item \textbf{Discrete placement} - the PCB is modeled as a main component surrounded by fixed candidate locations, each corresponding to a discrete action. Modeling the PCB as a continuous 2D plane contributes small meaningless shifts on the physical board, inflating the search space and rendering optimization intractable. Adopting a discrete action space provides a compact action space, reducing the RL search complexity while preserving placement feasibility.

\item \textbf{Net proximity} - we leverage prior knowledge that each passive must be placed near its associated voltage source. Net proximity information from the circuit schematic can be incorporated into the reward function, guiding the RL and avoiding wasted exploration of infeasible or physically impractical locations.
\end{itemize}
 
\begin{figure*}[t]
\centering

\subcaptionbox{SA}{
\includegraphics[scale=0.06]{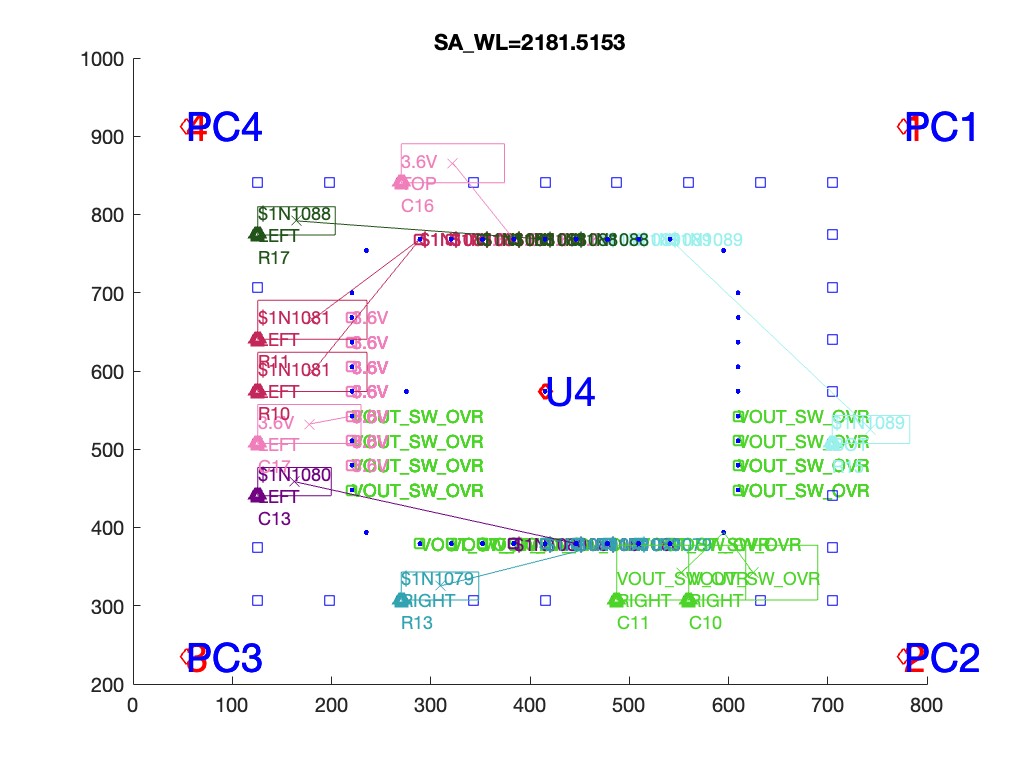}
\includegraphics[scale=0.06]{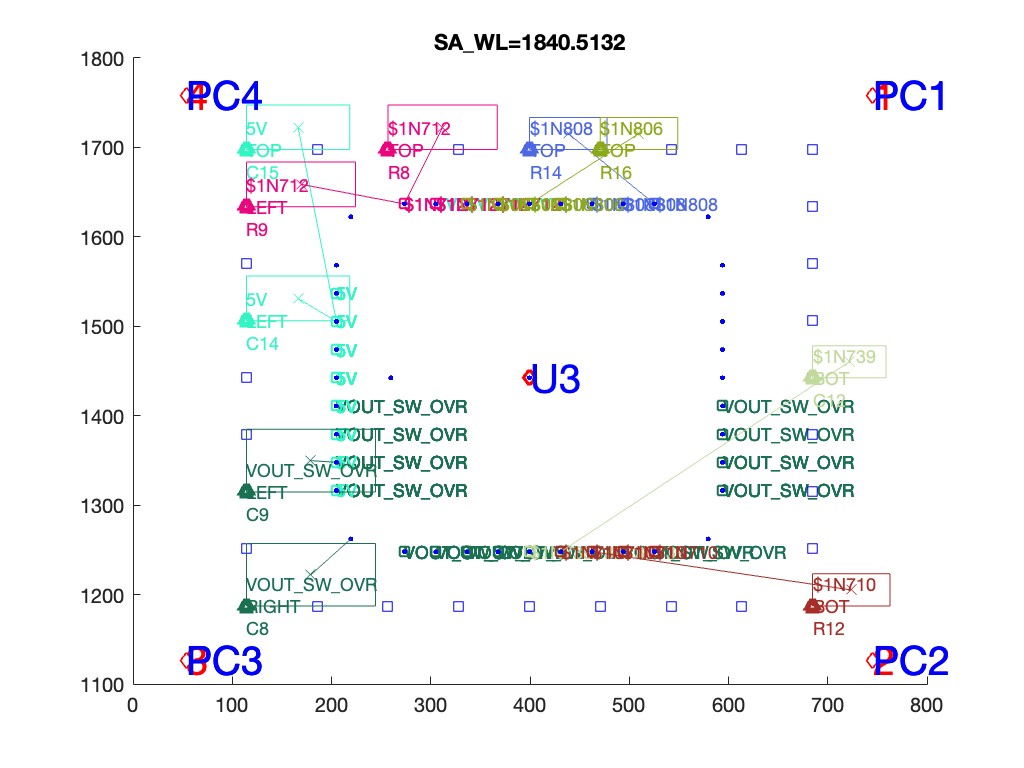}
\includegraphics[scale=0.06]{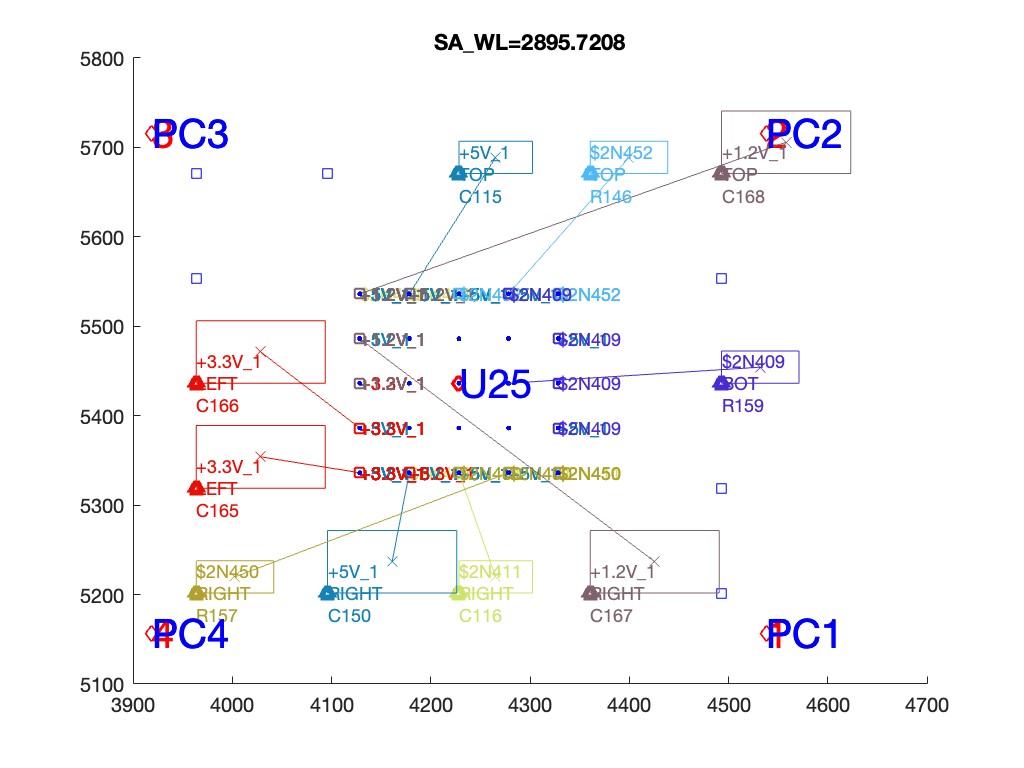}
\includegraphics[scale=0.06]{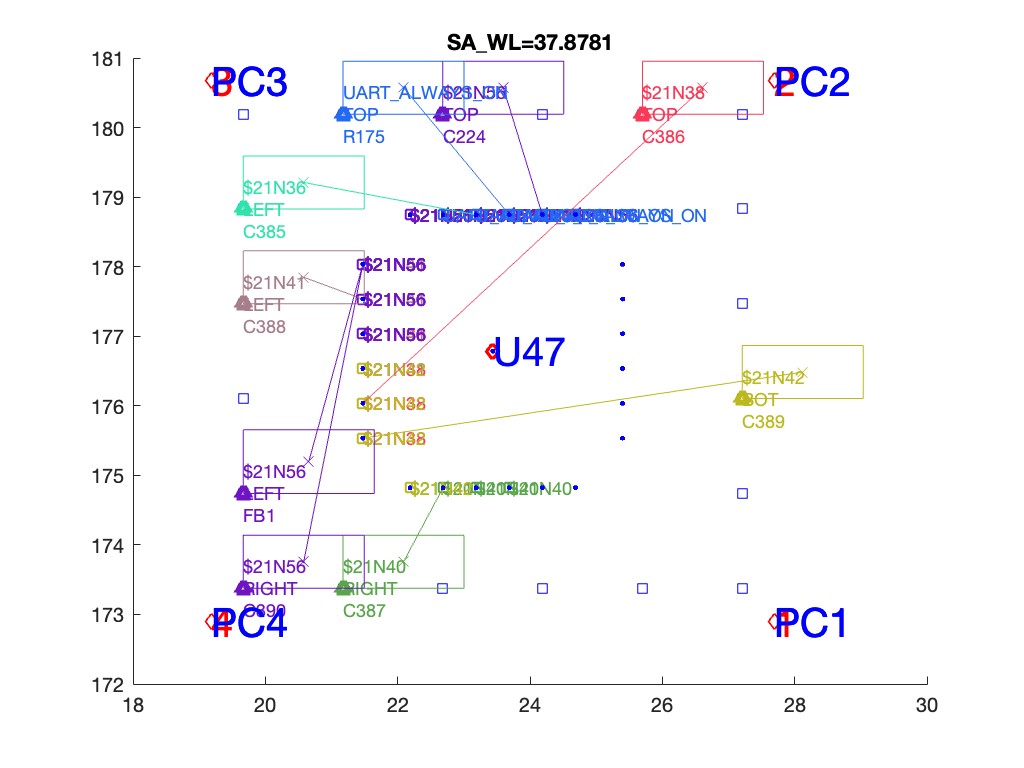}
\includegraphics[scale=0.05]{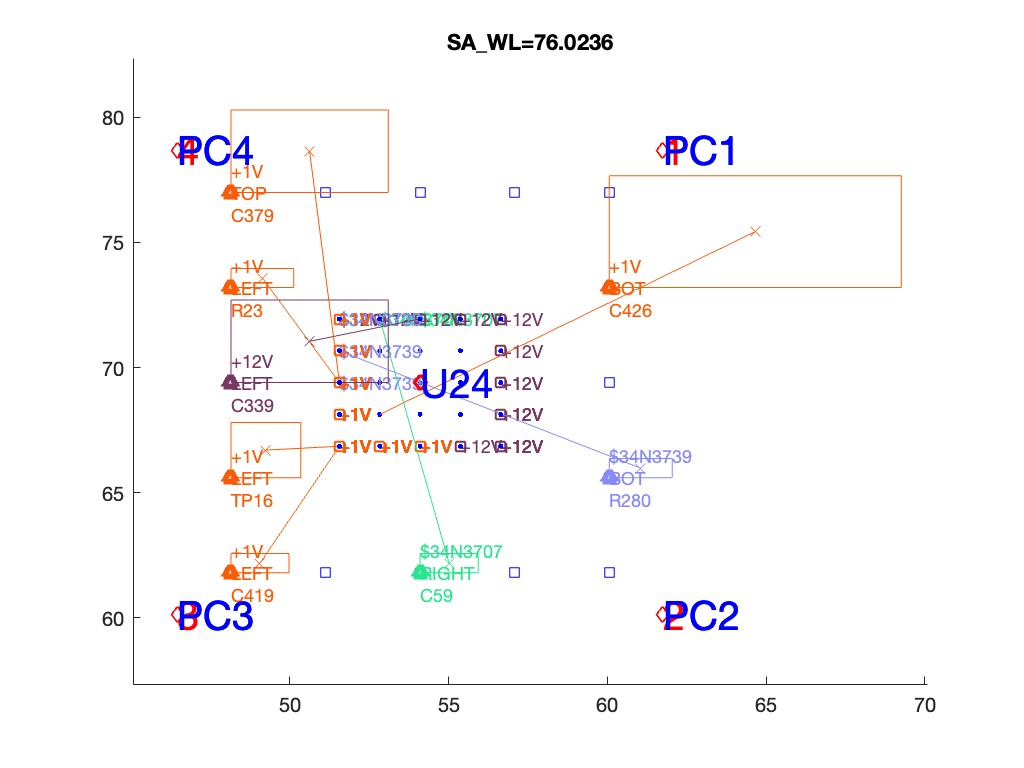}
\includegraphics[scale=0.06]{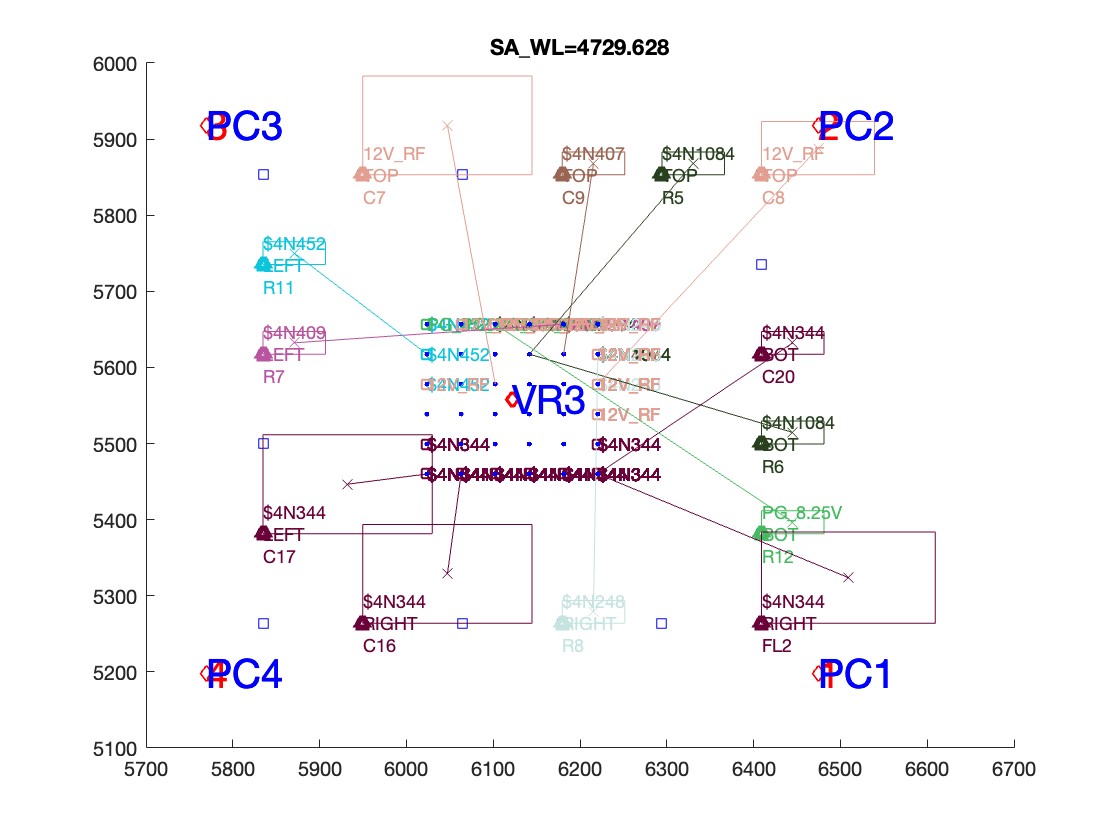}
\includegraphics[scale=0.06]{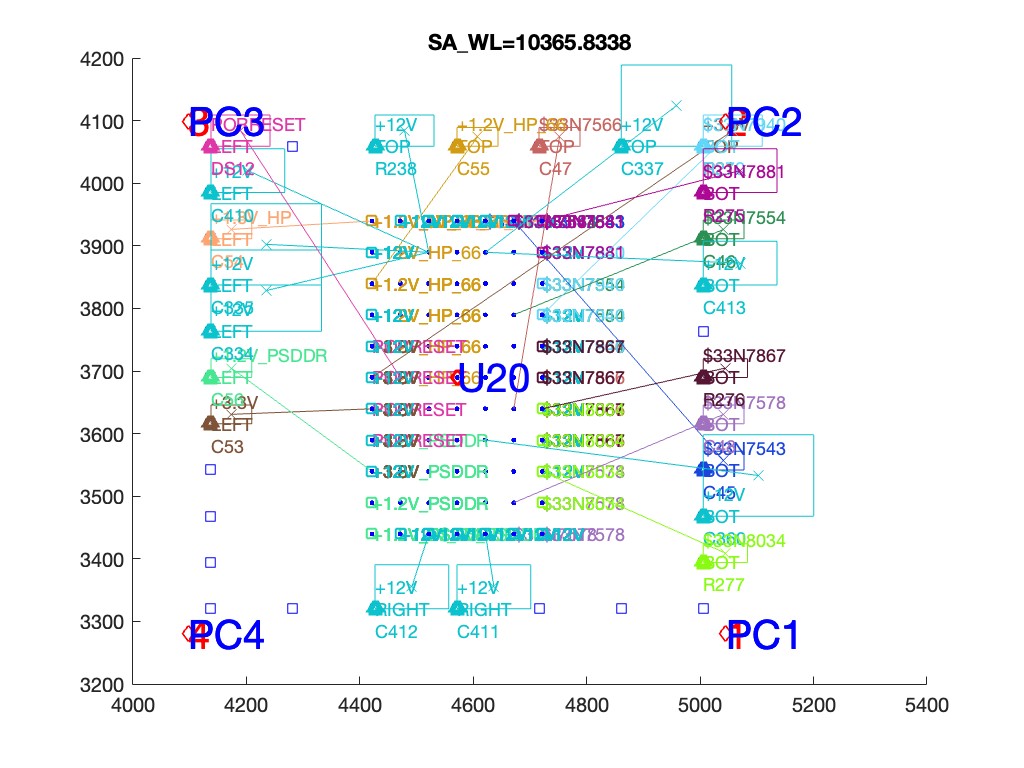}
\includegraphics[scale=0.06]{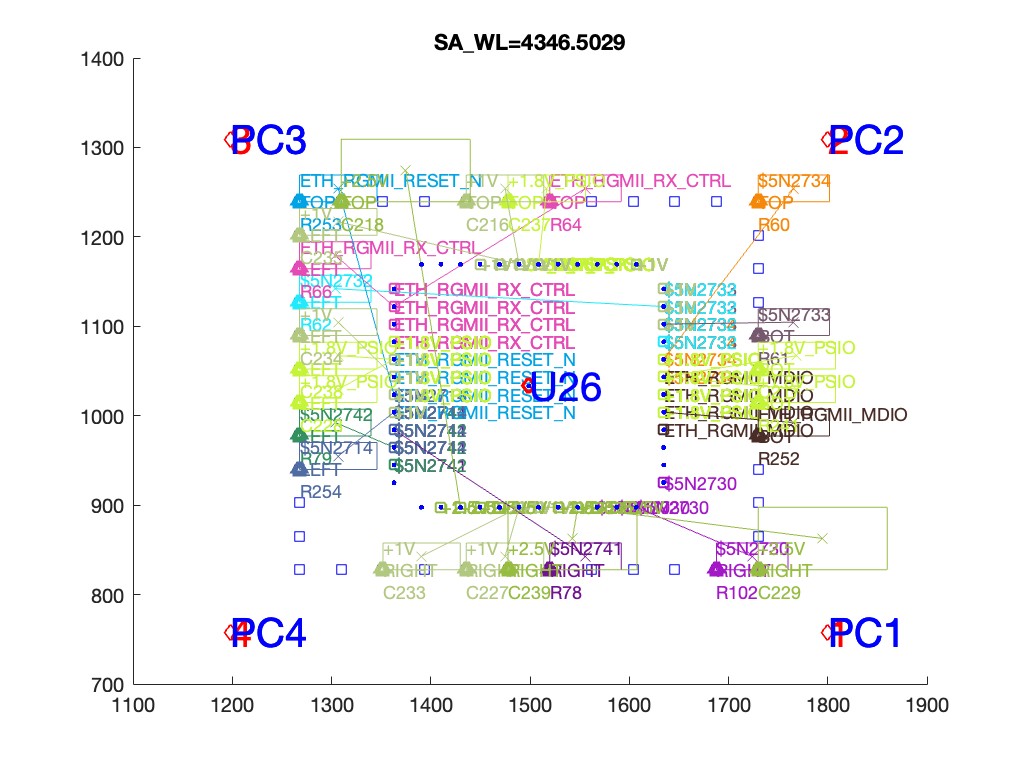}
}


\subcaptionbox{DQN}{
\includegraphics[scale=0.06]{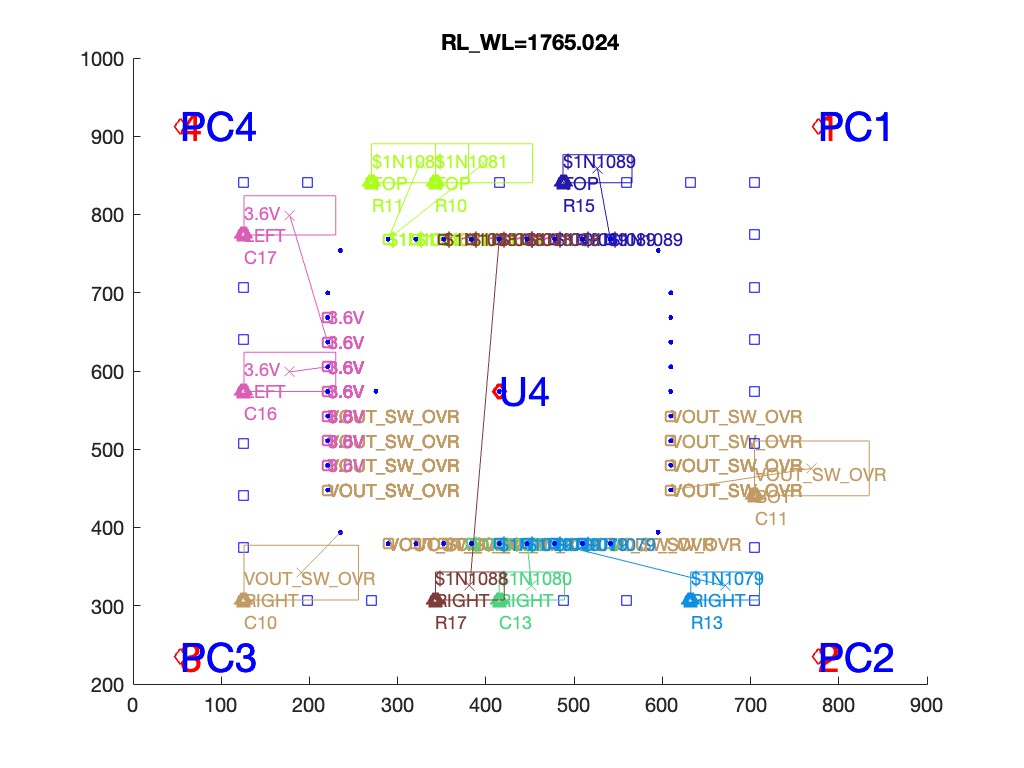}
\includegraphics[scale=0.06]{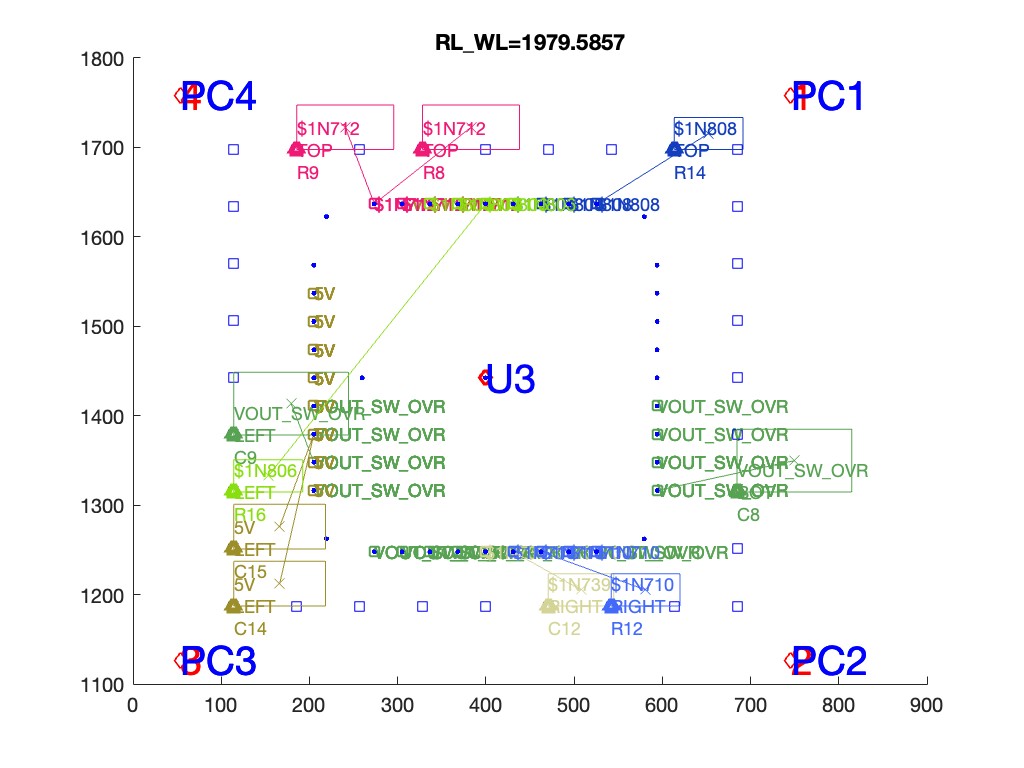}
\includegraphics[scale=0.06]{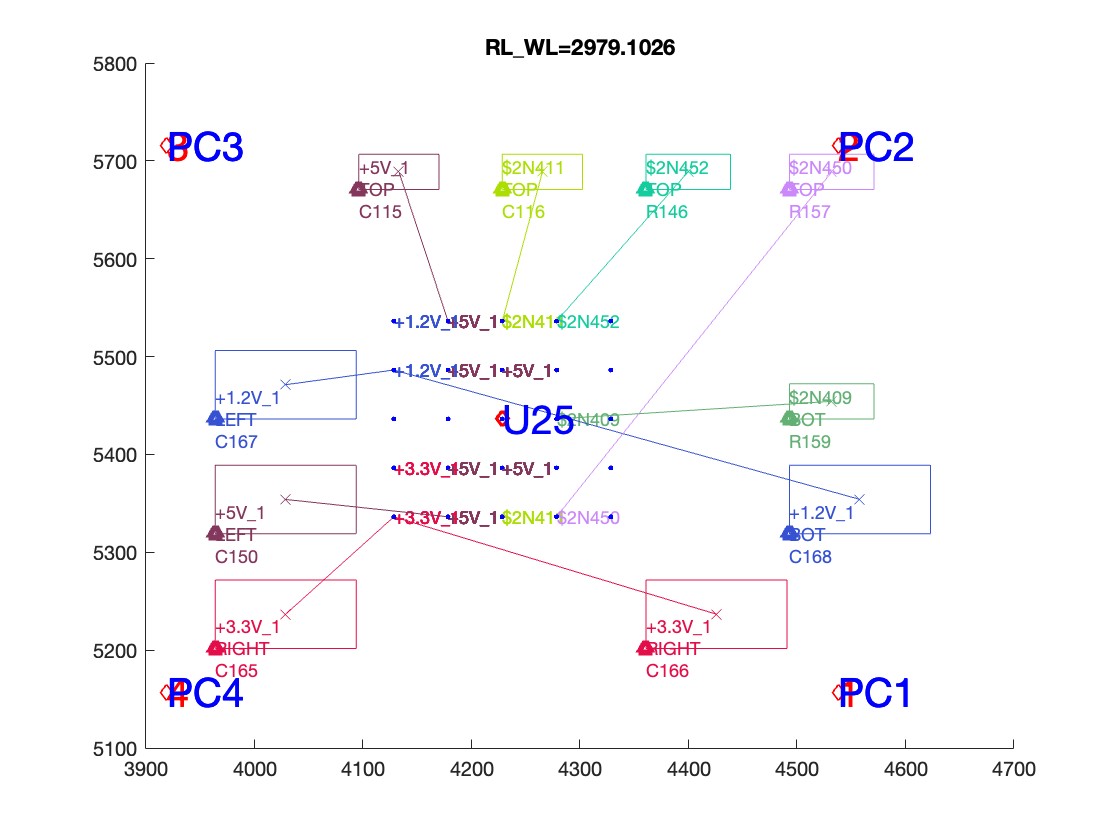}
\includegraphics[scale=0.06]{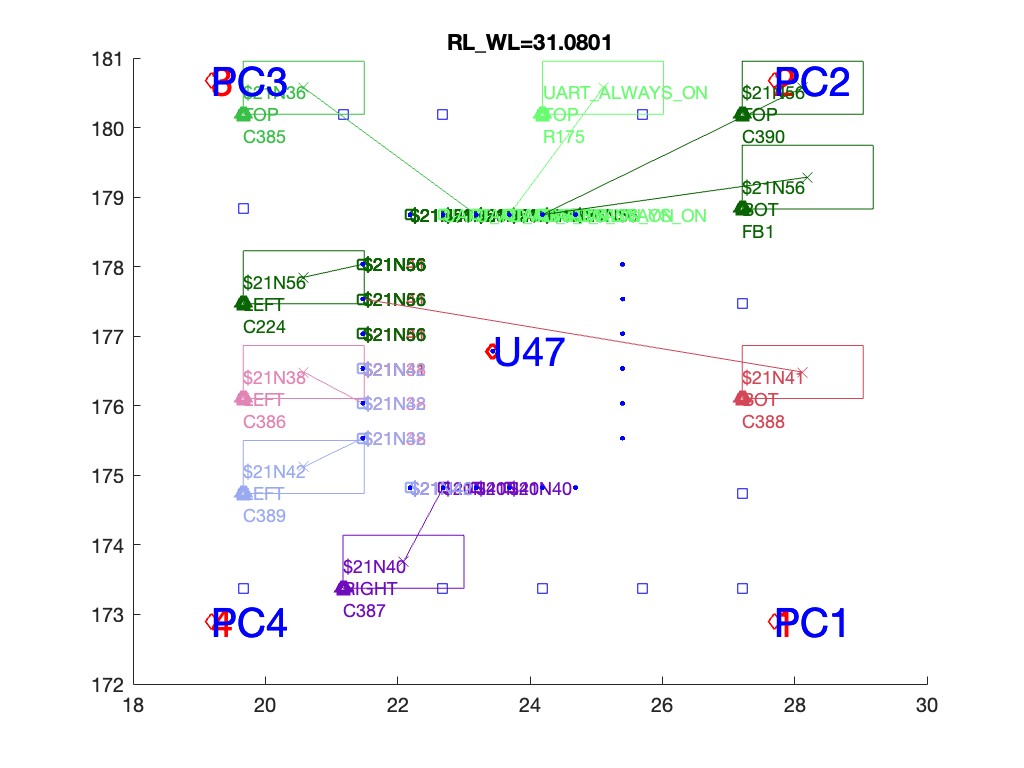}
\includegraphics[scale=0.05]{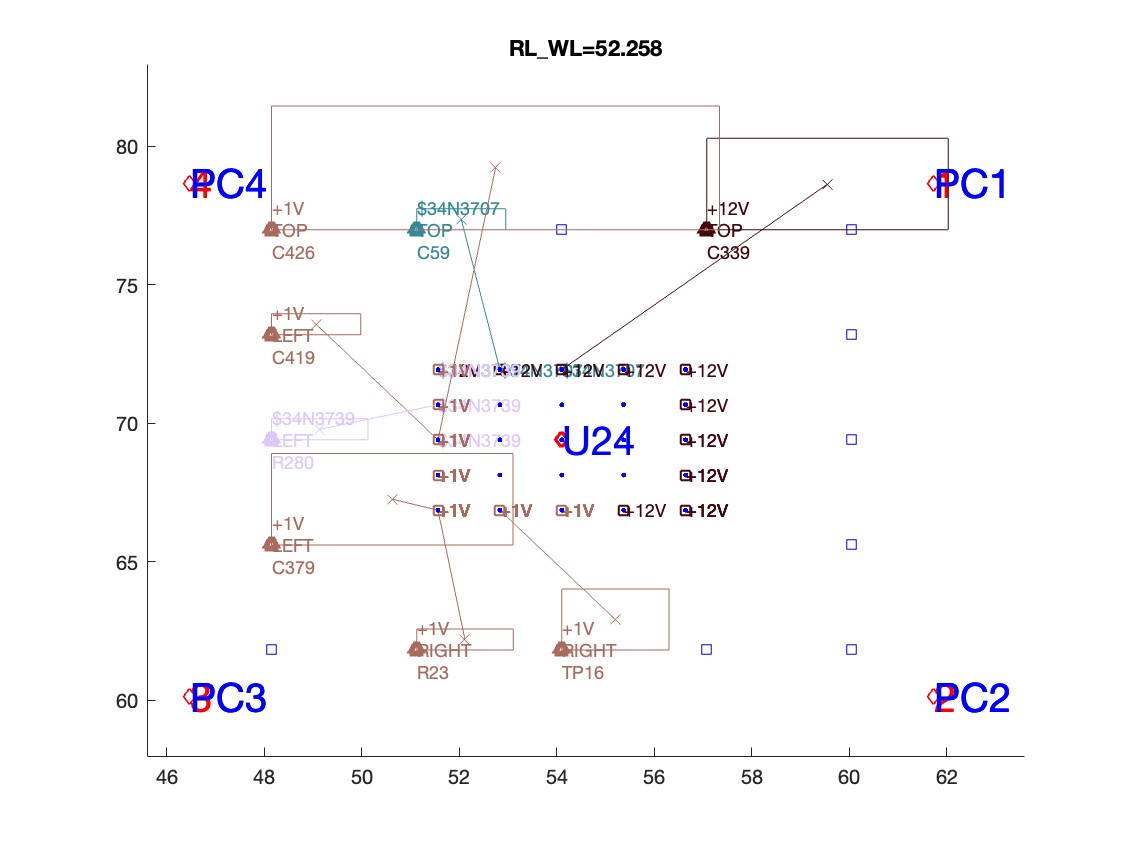}
\includegraphics[scale=0.06]{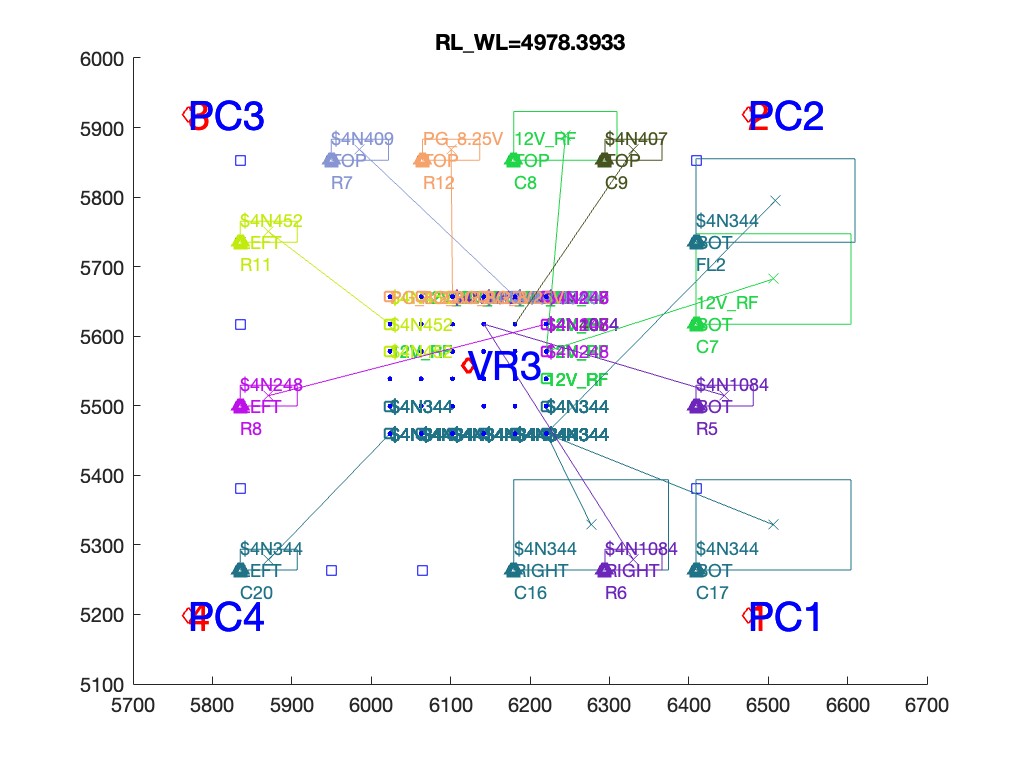}
\includegraphics[scale=0.06]{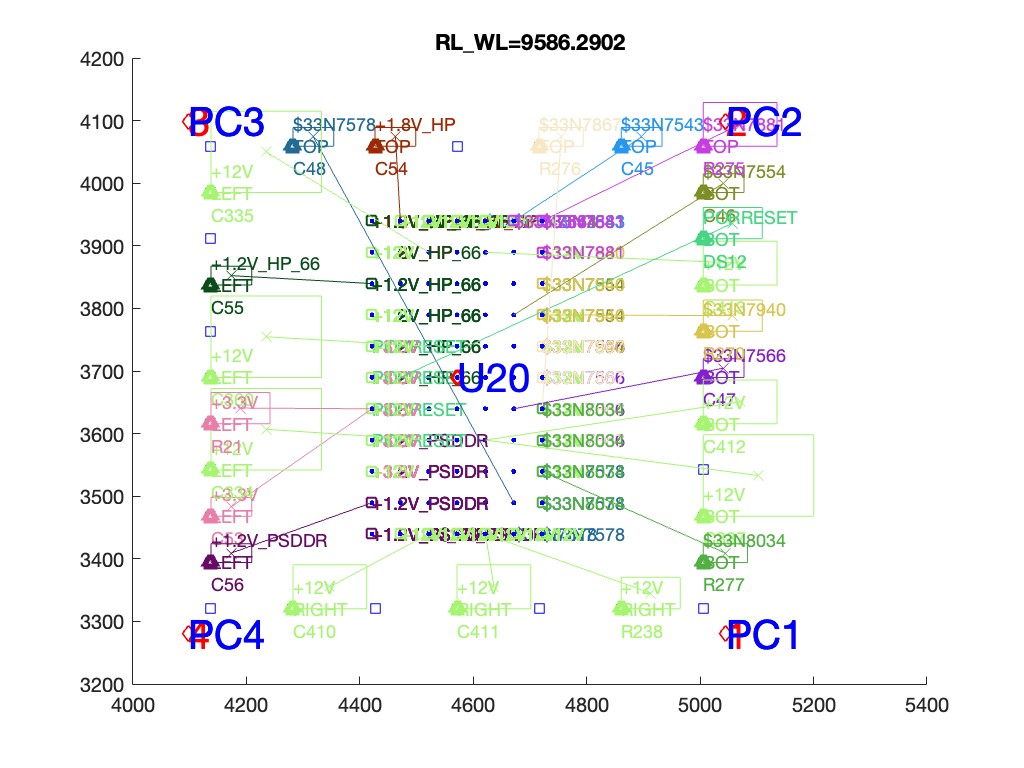}
\includegraphics[scale=0.06]{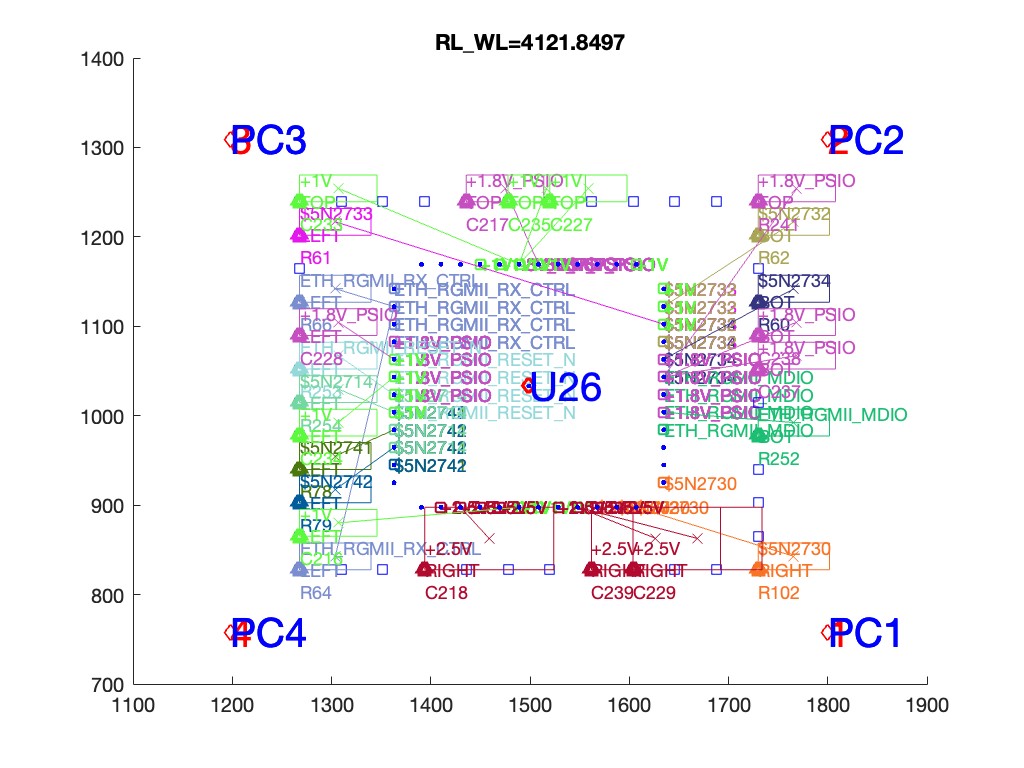}
}


\subcaptionbox{A2C}{
\includegraphics[scale=0.1]{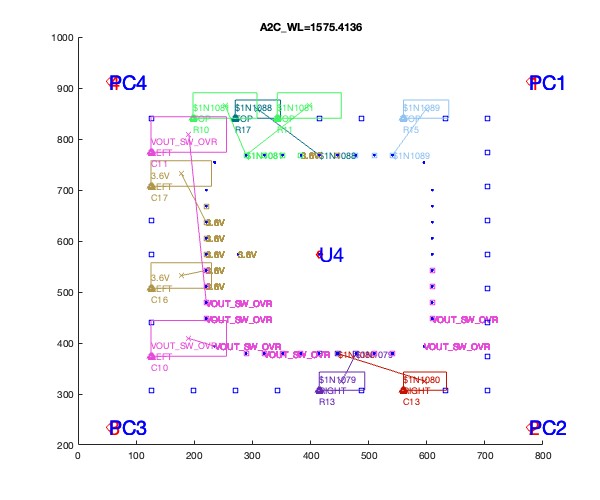}
\includegraphics[scale=0.1]{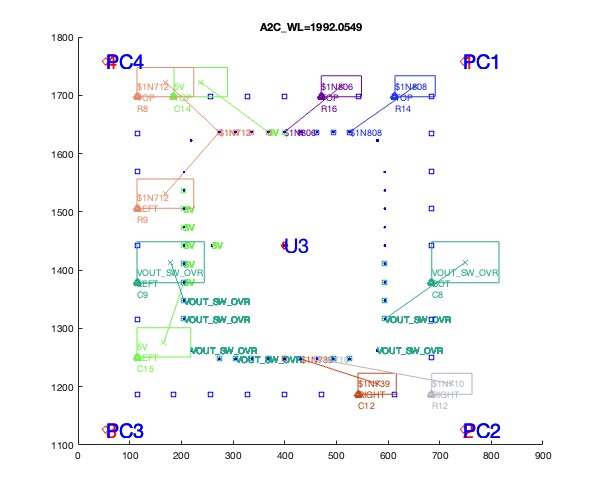}
\includegraphics[scale=0.1]{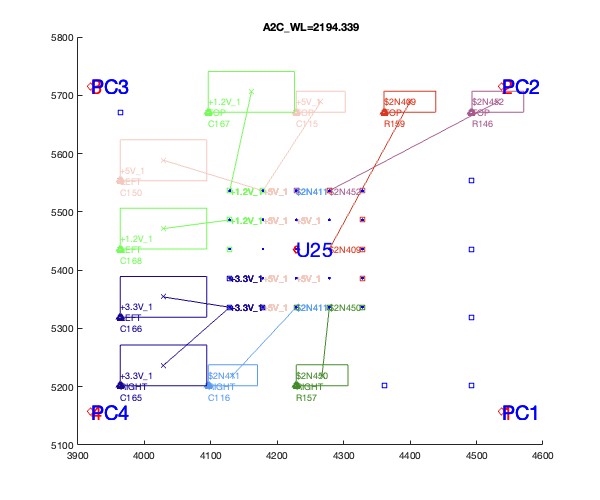}
\includegraphics[scale=0.11]{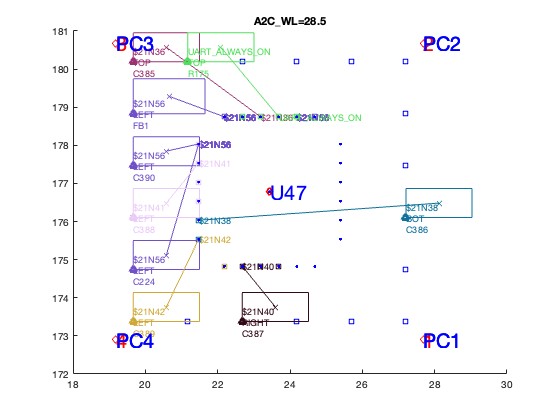}
\includegraphics[scale=0.1]{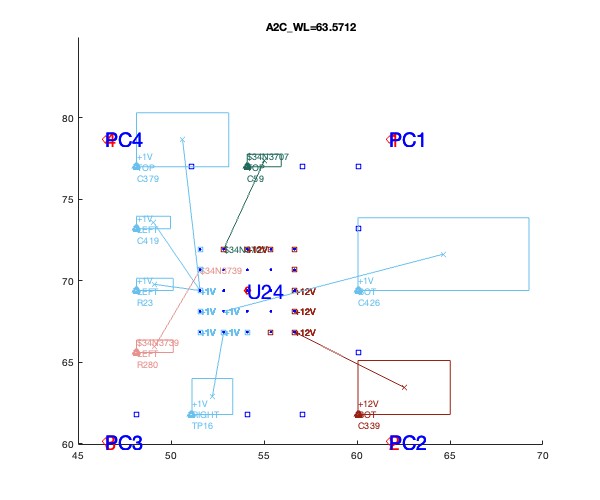}
\includegraphics[scale=0.11]{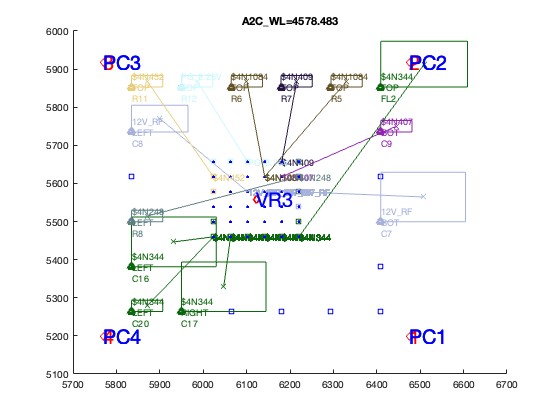}
\includegraphics[scale=0.11]{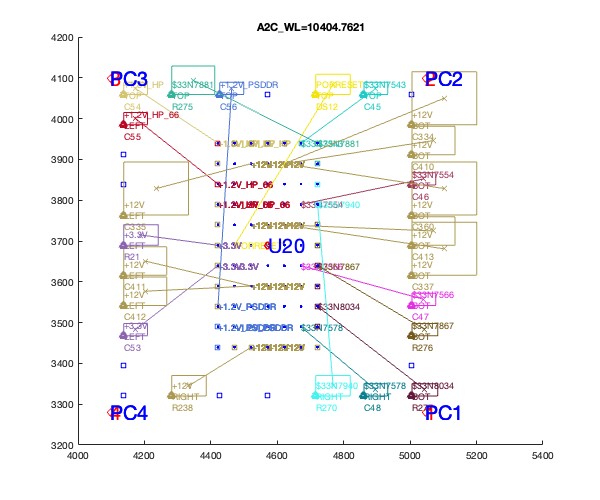}
\includegraphics[scale=0.06]{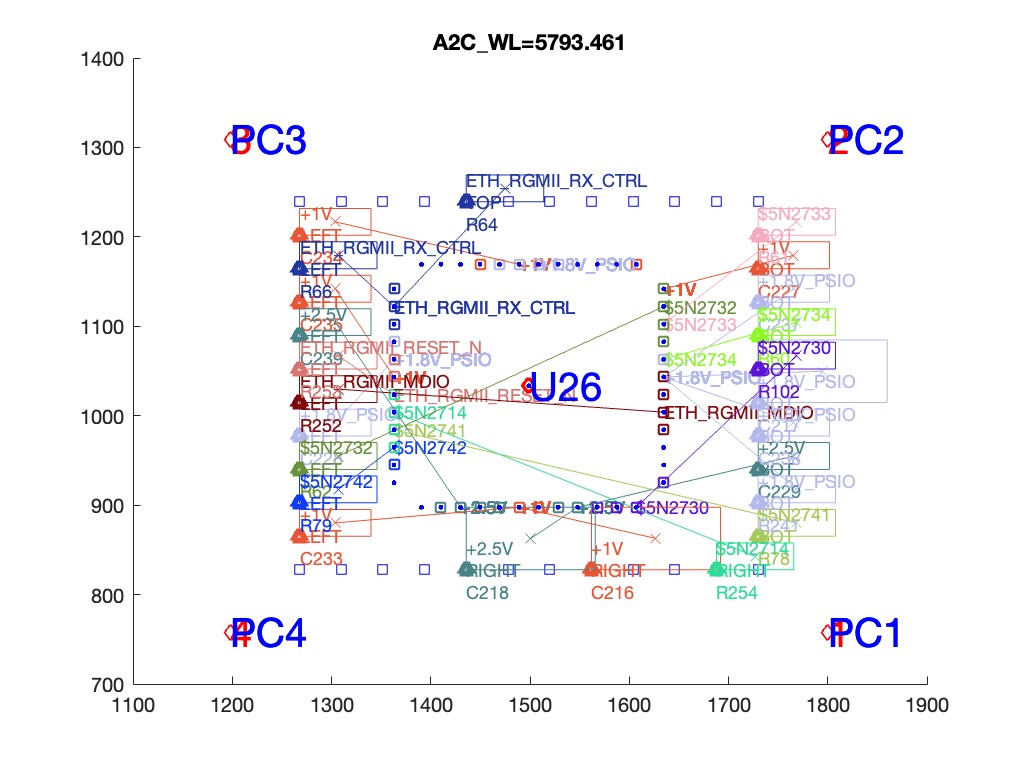}
}


\subcaptionbox{DQNnet}{
\includegraphics[scale=0.06]{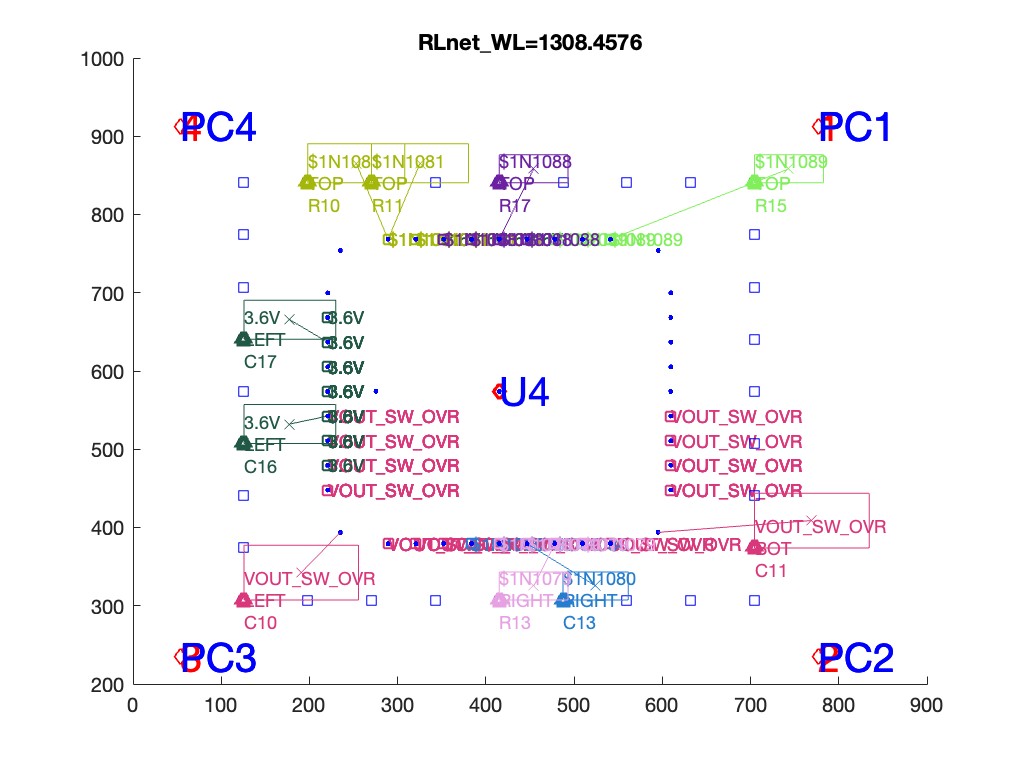}
\includegraphics[scale=0.06]{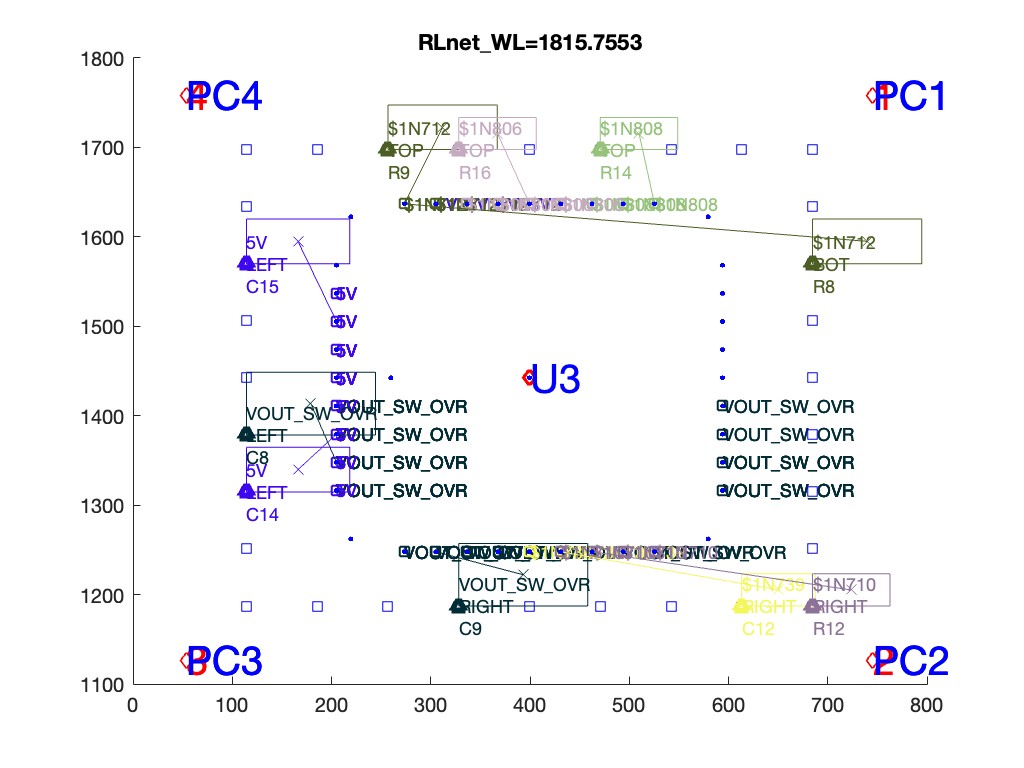}
\includegraphics[scale=0.06]{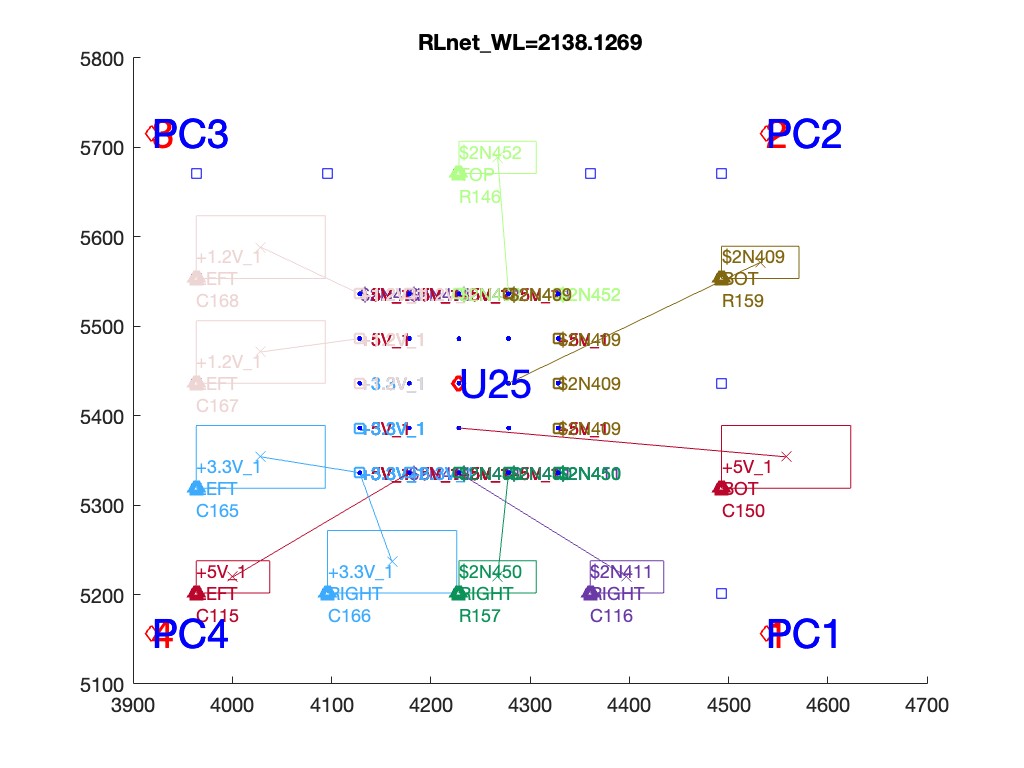}
\includegraphics[scale=0.06]{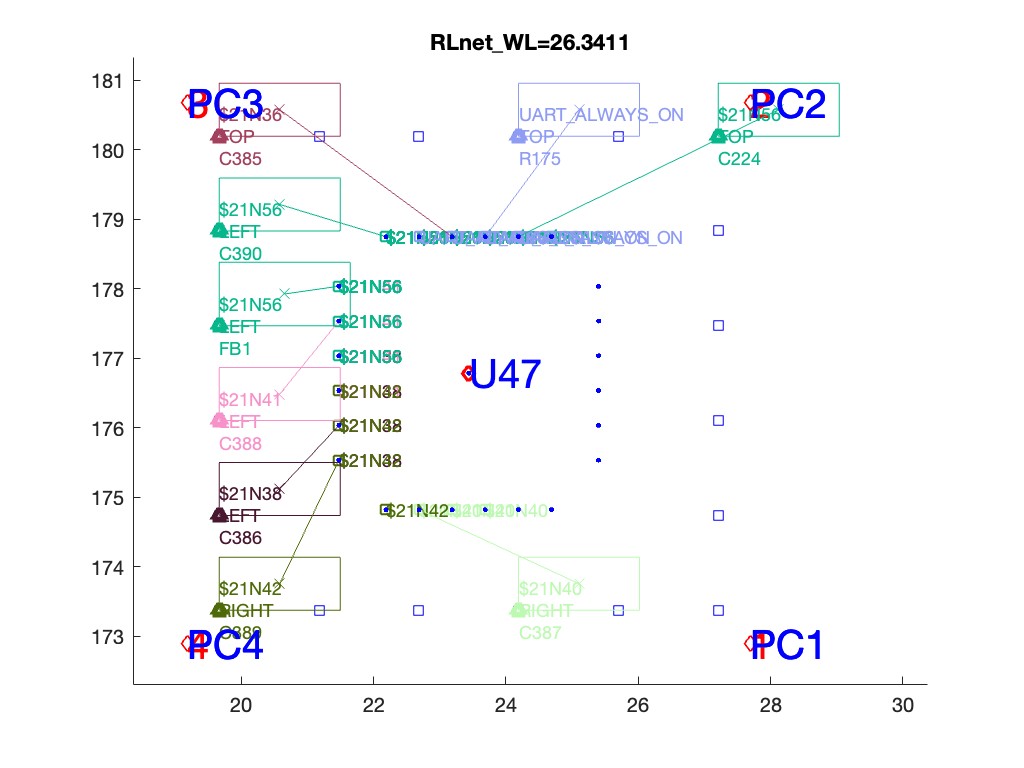}
\includegraphics[scale=0.05]{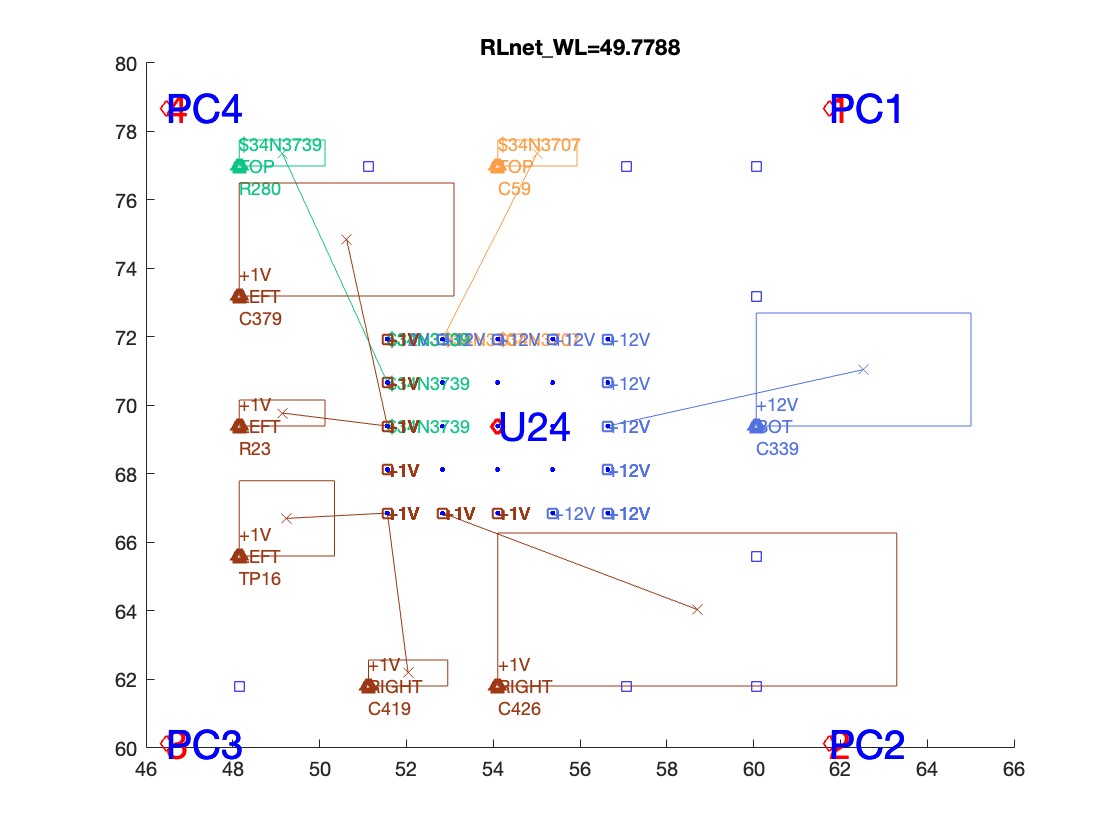}
\includegraphics[scale=0.06]{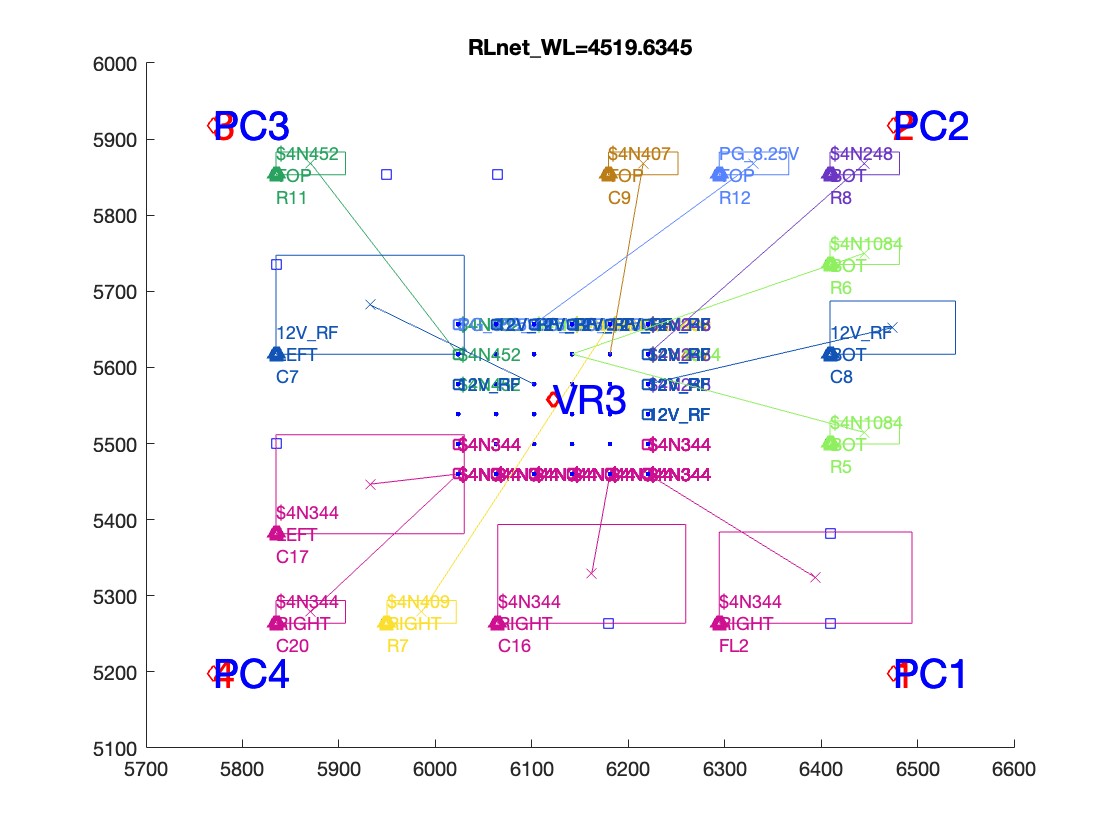}
\includegraphics[scale=0.06]{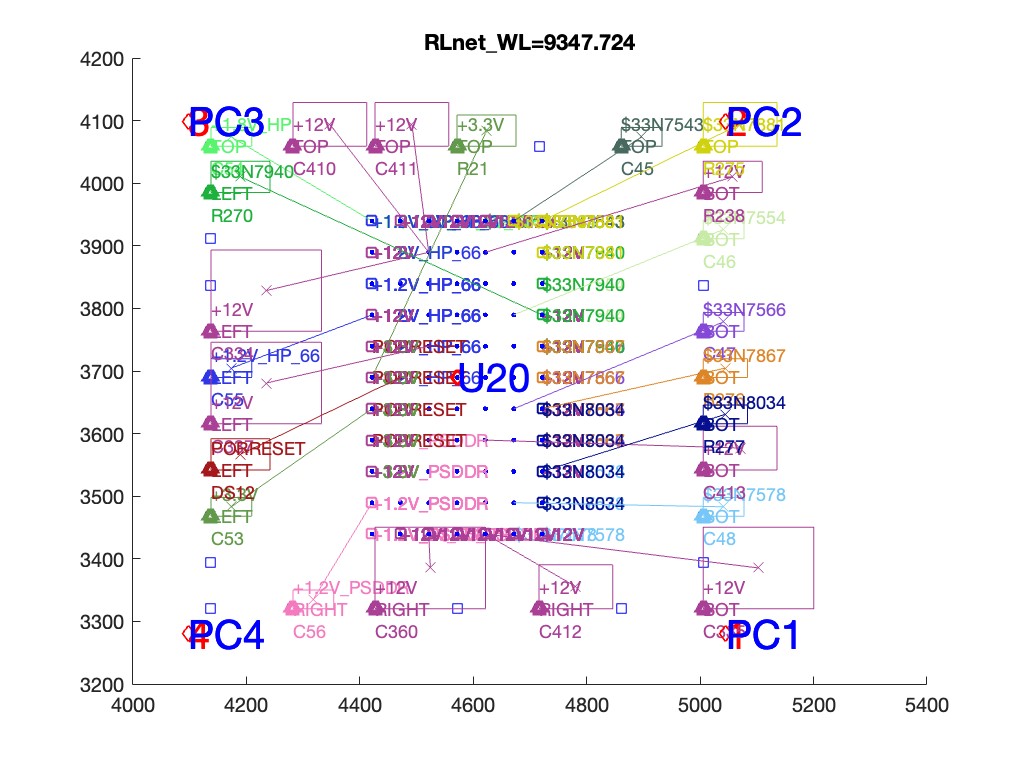}
\includegraphics[scale=0.06]{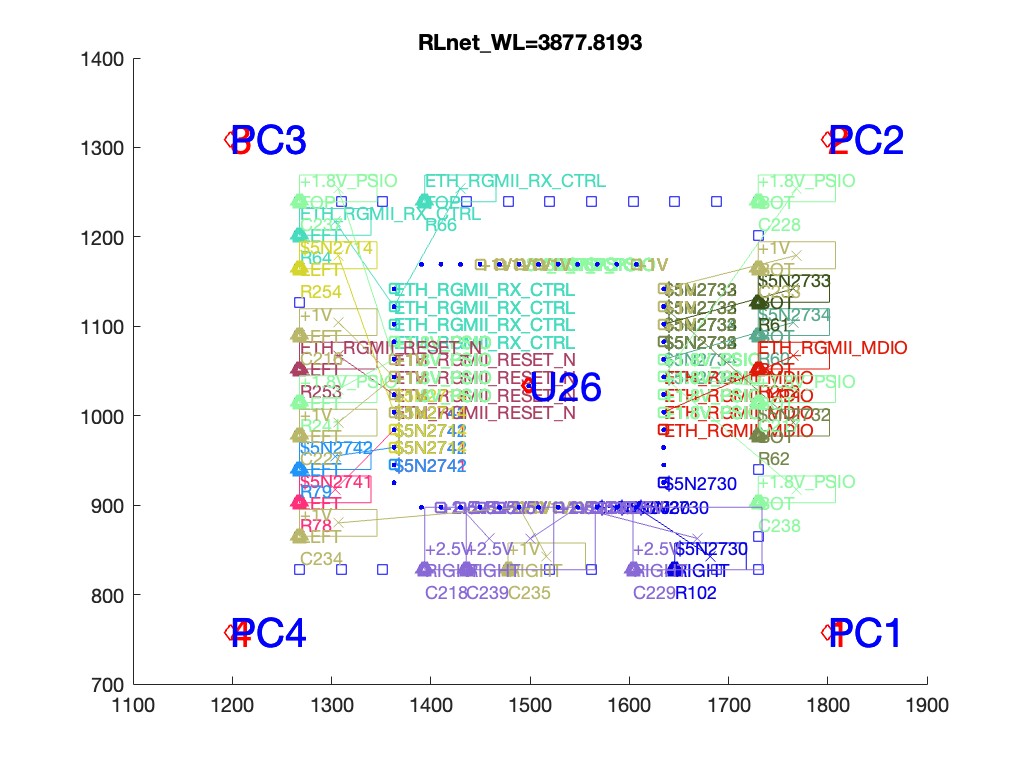}
}


\subcaptionbox{Ground truth}{
\includegraphics[scale=0.06]{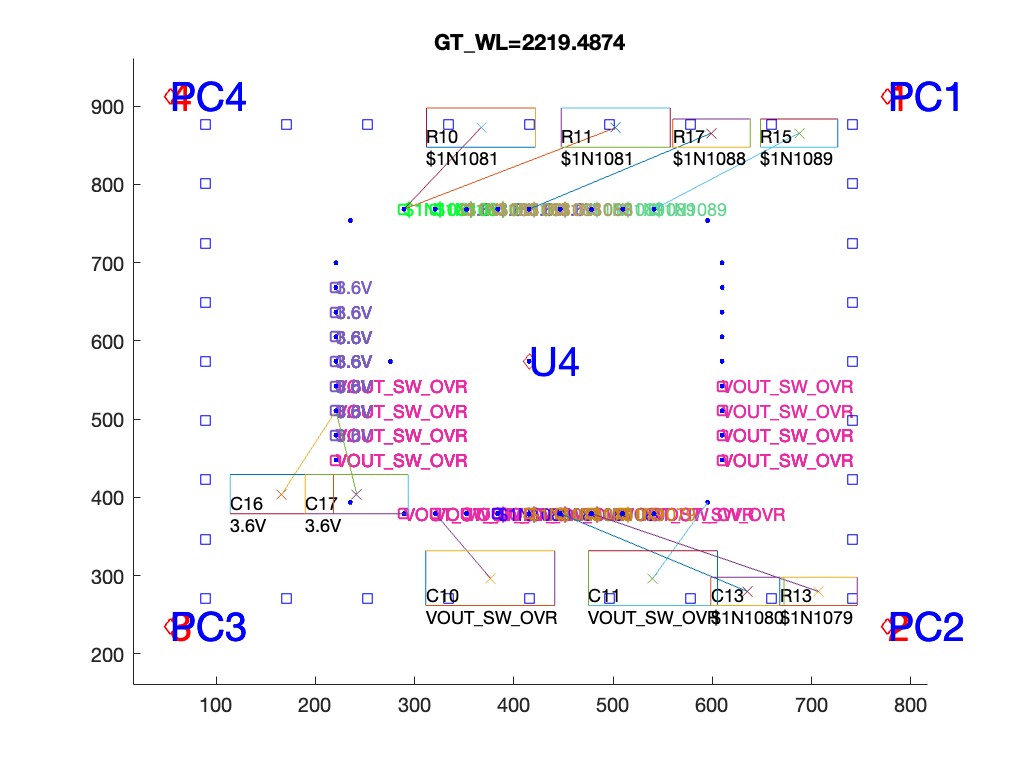}
\includegraphics[scale=0.06]{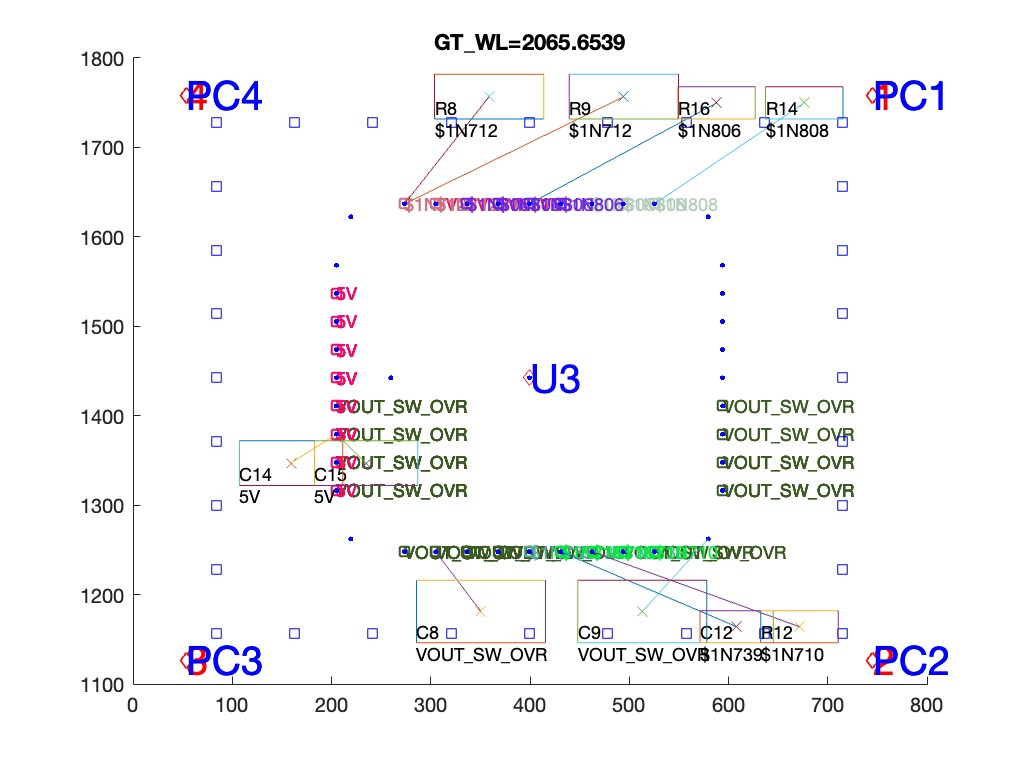}
\includegraphics[scale=0.06]{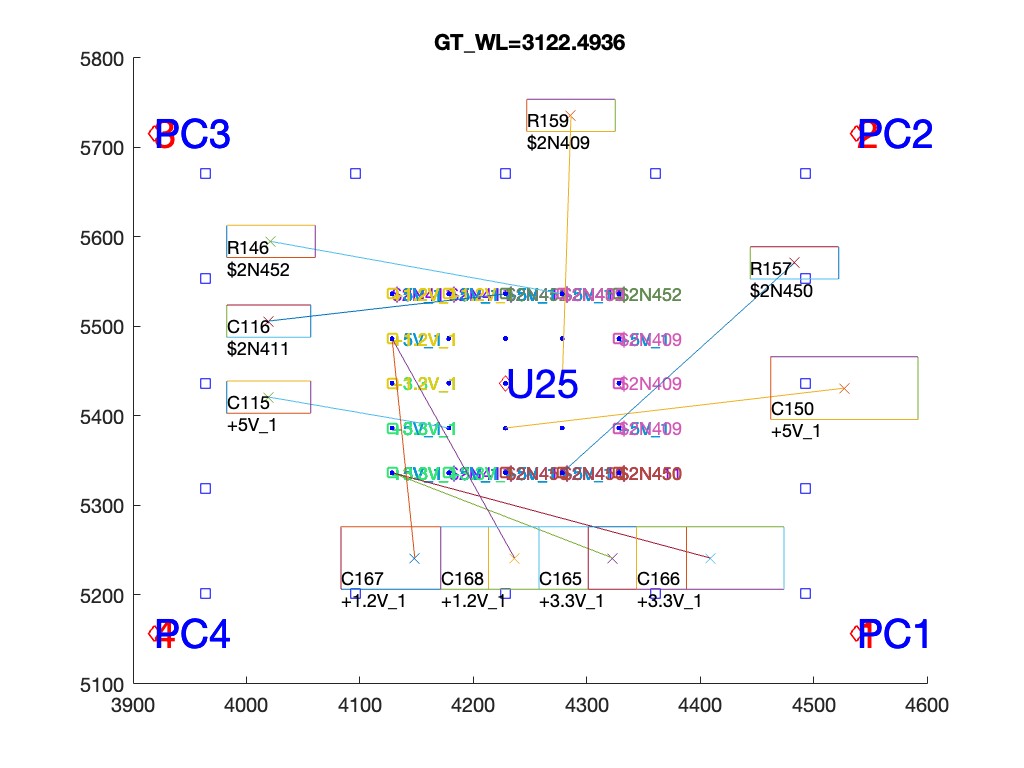}
\includegraphics[scale=0.06]{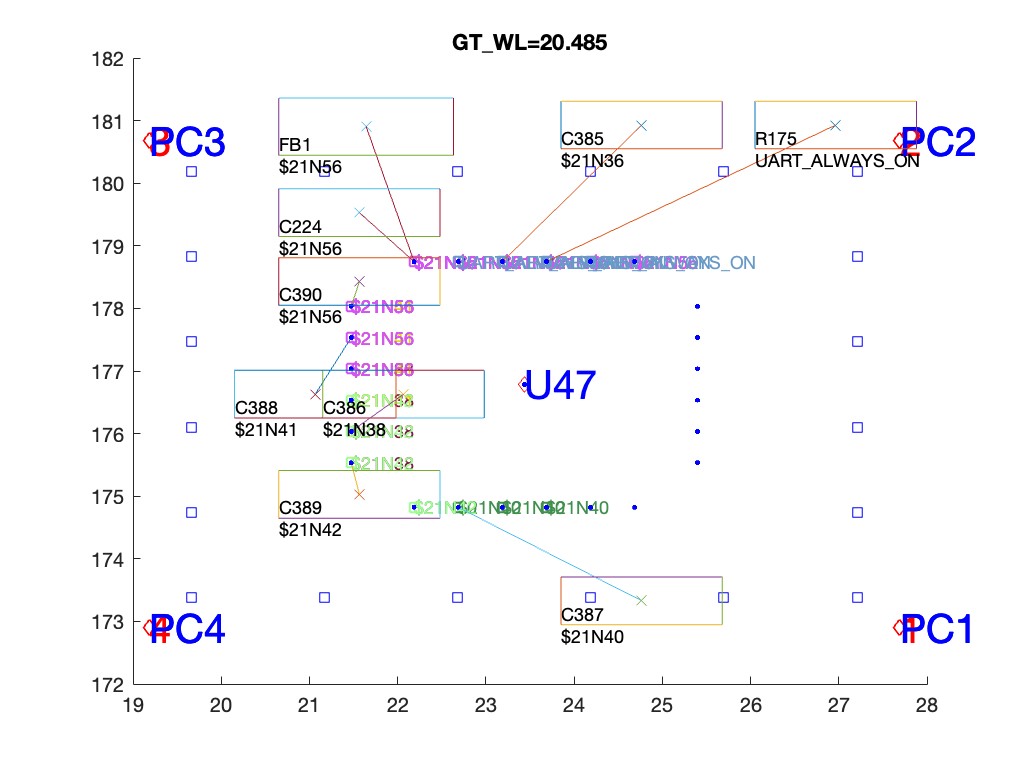}
\includegraphics[scale=0.06]{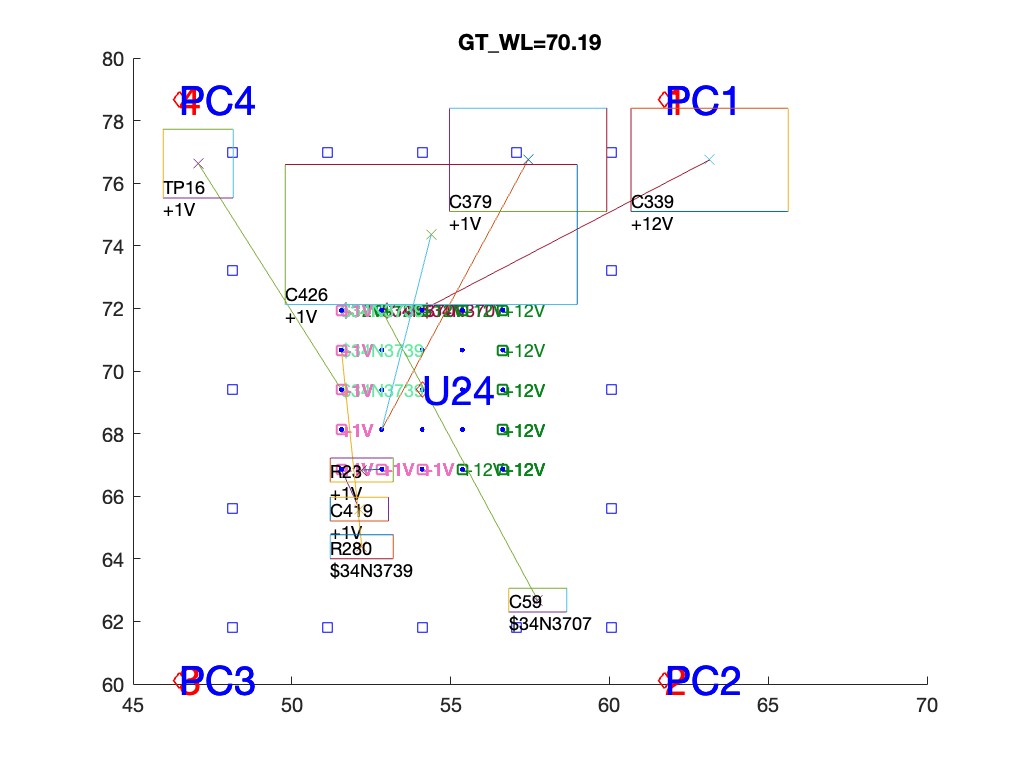}
\includegraphics[scale=0.06]{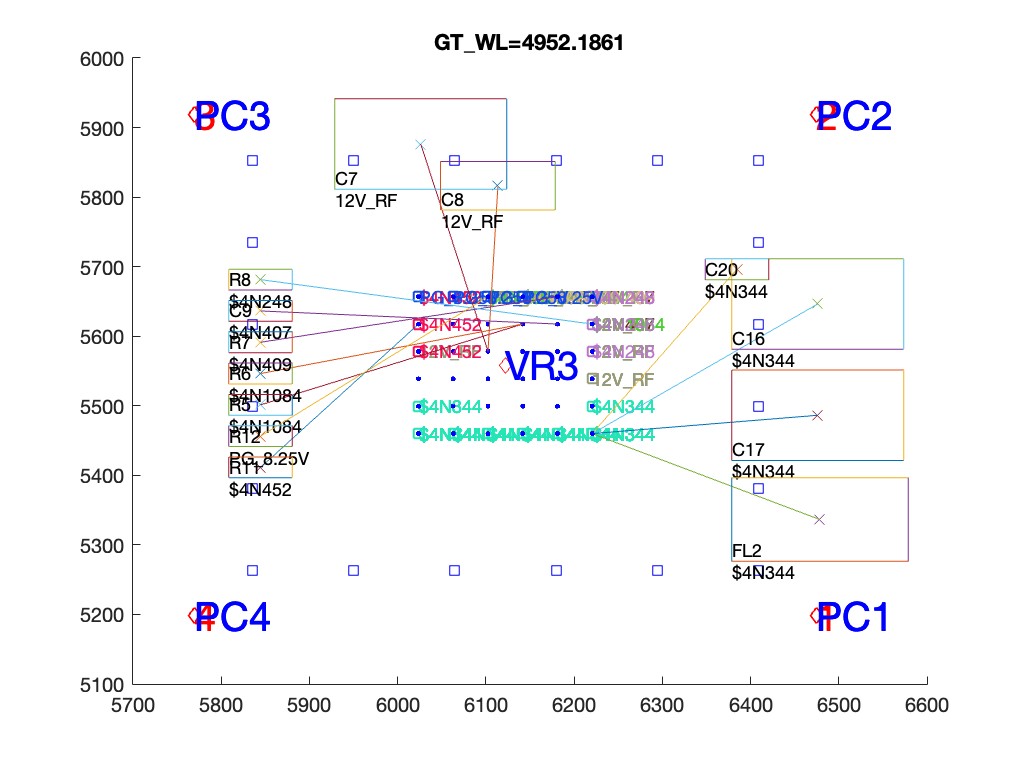}
\includegraphics[scale=0.06]{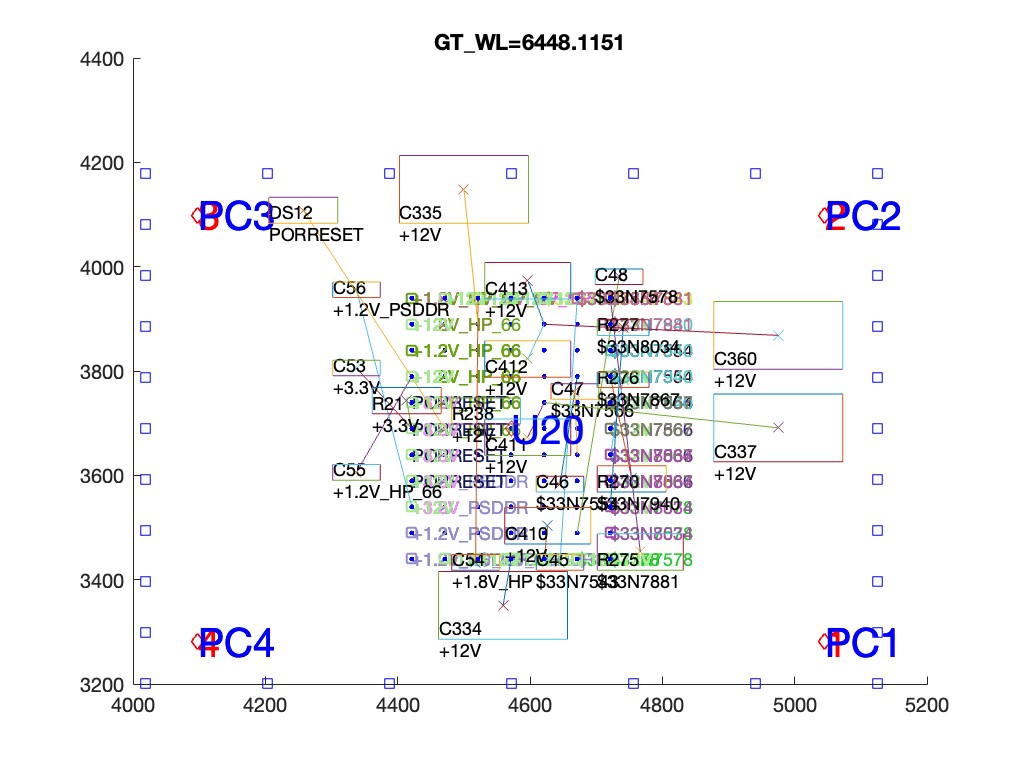}
\includegraphics[scale=0.06]{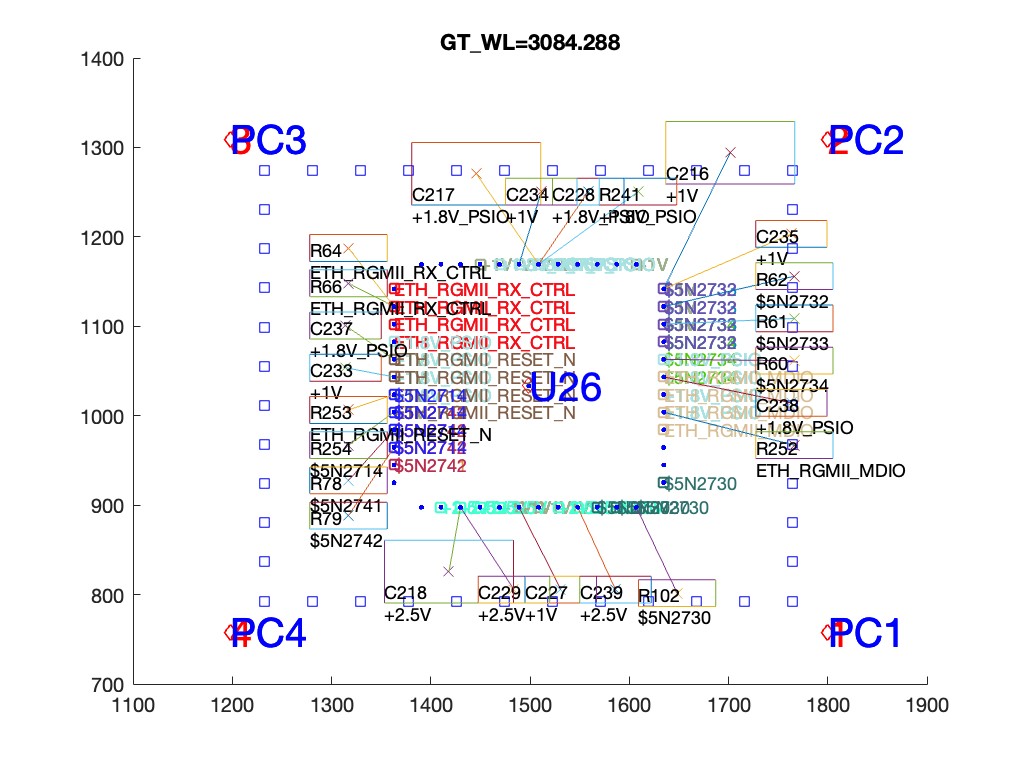}
}

\caption{Comparison of placement results across different methods}
\label{fig:groundtruth}

\end{figure*}

\subsection{Placement Overview}

We represent PCB placement as a set of $N$ number of discrete physical locations, $\mathcal{L}=\{l_{1},l_{2},\dots,l_{N}\}$, arranged around the border of a main component, where each $l_{i}=(x_{i},y_{i})\in\mathbb{R}^{2}$ is a valid placement location. In RL, action space is defined as $\mathcal{A}=\{a_{1},\dots,a_{N}\}$ where $a_{i}\in\left\{ 0,1\right\} $, $a_{i}=1$ places a component at $l_{i}$. The state $S=\{s_{1},s_{2},\dots,s_{M}\}$ where each $s_{i}$ is a one-hot vector of length $M$, indicating the current component ID to be placed. A state--action table, $sQp\in\mathbb{R}^{\left|S\right|\times\left|A\right|}$ is generated each episode either by exploration or exploitation conditions. Reward table, $\Gamma\in\mathbb{R}^{\left|S\right|\times\left|A\right|}$, captures power pins at the main component. But $\Gamma$ imposes a hard-constraint i.e. $a=a_{\text{netlist}}\Rightarrow\text{reward}>0,\quad a\neq a_{\text{netlist}}\Rightarrow\text{reward}=0$. Thus, we introduce $\text{Top-}K(s)$ to relax $\Gamma$ by defining $\text{Top-}K(s)\subseteq\mathcal{A},\quad|\text{Top-}K(s)|=K$. Also, $\text{Top-}K(s)$ refers to the $K$ nearest neighbors actions to centroid of $a_{netlist}$. All actions in $Top-K(s)$ receives $reward>0$, thereby expanding $\Gamma$ and facilitating exploration. Reward functions explained next section shows the interactions between $sQp$ and $\Gamma$. Each episode, RL maps states to actions, $\pi:\mathcal{S}\to\mathcal{A}$.

\subsection{Reward Functions}

Each candidate placement is evaluated using a weighted reward function, where the total reward $R_{total}$, is expressed as a weighted sum of overlap avoidance and net proximity. The weighting factor $\alpha\in[0,1]$ encourages placements that minimize overlaps between passive components, while the complementary term emphasizes net proximity, which directly reduces wirelength.

\begin{equation}
R_{total}=\alpha R_{non-overlap}+(1-\alpha)R_{proximity}
\label{eq:reward}
\end{equation}

\begin{itemize}
\item \textbf{Non-overlap reward}, $R_{non-overlap}$, where $d(s)$ is the dimension of the passive:
\begin{equation}
\begin{array}{c}
R_{non-overlap}(s,a)=1\left[\left\Vert o\left(sQp(s,a)\right),o\left(sQp(s,a')\right)\right\Vert _{2}\right]\\
>d(s),\;a'\neq a
\end{array}
\label{eq:nonoverlap}
\end{equation}

\item \textbf{Net proximity reward}, $R_{proximity}$, using $Qp$ and $\Gamma$ under constraints:
\begin{equation}
R_{proximity}(s,a)=1\left[sQp(s,a)=\Gamma(s,a)\right]
\label{eq:proximity}
\end{equation}
\end{itemize}

\subsection{RL Training}

The DQN loss function computed in Equation \ref{eq:dqn_loss} is using the outputs of two neural networks: the predictor network $Qp$, and target network $Qt$ evaluated at each state. $R_{total}$ is the observed reward, $\gamma$ is the discount factor. This loss is used to update the weights of $Qp$, while $Qt$ is updated more slowly to stabilize training. To incorporate policy learning, we can combine DQN-style critic \cite{mnih2016asynchronous} with an Actor loss based on policy gradient, giving the Actor-Critic loss in Equation \ref{eq:ac_loss}.

\begin{equation}
L_{DQN}=E\left[\left(R_{total}+\gamma\max_{a'}Q_{t}(s',a')-Q_{p}(s,a)\right)^{2}\right]
\label{eq:dqn_loss}
\end{equation}

\begin{equation}
L_{\text{AC}}=-\log\pi_{\theta}(a\mid s)\,Q_{p}(s,a)+L_{\text{DQN}}
\label{eq:ac_loss}
\end{equation}

\subsection{Wirelength}

Past studies on physical design have commonly used Half Perimeter Wirelength (HPWL) as one of the learning objective \cite{mirhoseini2020chip,cheng2021joint}. New studies \cite{vassallo2024learning,chen2025pcbagent}, including this work, adopts Total Euclidean Wirelength (TEWL). The motivation is that HPWL only depends on the extremes that contains all pins of the netlist, so adding additional pins within the bounding box does not change its value. In contrast, TEWL accounts for the actual distances between all pins, providing a better correlation with the routed wirelength. Formally, we define TEWL in Equation \ref{eq:tewl}, where $\{d,o,\mathcal{N}\}$ refers to the state $s$ (or passive)'s dimension (e.g. ABS Length and Breadth) , the predicted placement location of the passive, pin location of the main component that share the same netname as the passive (different pins $t$ at different locations may share the same netname) respectively. The optimal action $a^{*}$ is predicted based on the chosen RL method: DQN or actor--critic (AC).

\begin{equation}
\begin{array}{c}
\text{TEWL}=\sum_{\text{s}=1}^{finalstate}\;\underset{t}{\min}\;\Big\|\frac{1}{2}\mathbf{d}(s)+\mathbf{o}(a^{*}(s))-\mathbf{\mathcal{N}}(s,t)\Big\|_{1}\\
\\\text{DQN:}\quad a^{*}(s)=\underset{a\in\mathcal{A}}{\arg\max}\;Q_{p}(s,a)\\
\\\text{AC:}\quad a^{*}(s)=\underset{a\in\mathcal{A}}{\arg\max}\;\pi_{\theta}(a\mid s)
\end{array}
\label{eq:tewl}
\end{equation}

\subsection{Token Based Input}

In contrast to feature-based inputs \cite{vassallo2024learning}, where states are described using continuous attributes such as component coordinates, distances, or rotation, we represents passive ID as state, $s=p_{state}$, and uses a one-hot vector encoding. RL may use immediate reward or n-step reward. For immediate reward the learning is more stable. However, this means that each passive is treated as an independent placement in the PCB, losing long term reward \cite{mirhoseini2020chip}. On the other hand n-step reward is more expensive and can be unstable to learn. As a workaround, we use a token based input, by incorporate both passive and net ID as a single state. Intuitively, this works better because in PCB layouts, passives connected to the same net are usually placed in close physical proximity. We use one-hot vector to encode both passive and net information before concatenating them into a vector in Equation \ref{eq:state}.

\begin{equation}
\begin{array}{c}
s=[p_{state}\parallel n_{state}]\\
\\p_{\text{state}}\in\{0,1\}^{M},\quad n_{\text{state}}\in\{0,1\}^{N},\quad s\in\{0,1\}^{M+N}
\end{array}
\label{eq:state}
\end{equation}

\begin{table}
\caption{Dataset}
\label{tab:dataset}
\begin{center}
\begin{tabular}{|l|c|c|c|c|c|}
\hline
 & \#Passive & \#Nets & \#Action & \%ComDiff & GT \\
\hline
U4 & 10 & 7 & 36 & 29.2 & 2219 \\
U3 & 10 & 7 & 36 & 29.2 & 2065 \\
U25 & 10 & 7 & 20 & 29.2 & 3122 \\
U47 & 9 & 7 & 24 & 76.9 & 20 \\
U24 & 8 & 4 & 20 & 3.3 & 70 \\
U115 & 11 & 5 & 42 & 4.4 & 2564 \\
VR3 & 13 & 8 & 24 & 8.5 & 4952 \\
U20 & 23 & 14 & 36 & 8.5 & 6448 \\
U26 & 24 & 13 & 48 & 23.7 & 3084 \\
\hline
\end{tabular}
\end{center}
\end{table}

\begin{table}[htbp]
\centering
\caption{Wirelength (TEWL) Comparisons}
\label{tab:results}
\begin{tabular}{|l|c|c|c|c|}
\hline
 & \multicolumn{3}{c|}{Passive} 
 & Passive+Net\\
\cline{2-5}
 & SA & DQN & A2C & DQNnet \\
\hline
U4   & 2181 & 1765 & 1575 & \textbf{1308} \\
U3   & 1840 & 1979 & 1992 & \textbf{1815} \\
U25  & 2895 & 2979 & 2194 & \textbf{2138} \\
U47  & 37 & 31 & 28 & \textbf{26} \\
U24  & 76 & 52 & 63 & \textbf{49} \\
U115 & \textbf{2135} & 2666 & - & 2989 \\
VR3  & 4729 & 4978 & 4578 & \textbf{4519} \\
U20  & 10365 & 9586 & 10404 & \textbf{9347} \\
U26  & 4346 & 4121 & 5793 & \textbf{3877} \\
\hline
\end{tabular}
\end{table}

\begin{table}
\caption{Visual Inspection of MLs}
\label{tab:inspection}
\begin{center}
\begin{tabular}{|l|c|c|c|c|c|}
\hline
In Total PCBs & SA & DQN & A2C & DQNnet \\
\hline
Overlapping Passives & 14 & 12 & \textbf{4} & 6 
\\
Routing Conflicts & 22 & \textbf{13} & 16 & 16
\\
\hline
\end{tabular}
\end{center}
\end{table}

\section{Experiments}

\subsection{PCB Dataset and MLs}
 
We use an in-house dataset of 9 PCBs, representing power and digital functional groups, which were extracted from larger, more complex PCBs. We show in Table 1, the number of passive components (\#P) and number of nets (\#N), number of actions (\#A), component size disparities measured as the ratio between the smallest and largest passive (\%CD) and TEWL of human placement (GT). We rank the PCBs according to their complexity in Table 1 and GT in Fig 1. The first four PCBs are relative simple, each providing sufficient actions for placement. In terms of placement challenges, U24, U115, VR3 and U20 suffer from large disparity in component sizes, with U24 exhibiting the most pronounced disparity. Both U20 and U26 exhibit overlapping challenges, stemming from the large number of passives and the limited actions available to maneuver them. Also, U20 was originally designed by human for double-side placement, but we constrained it to a single-side configuration. 

The MLs used in this study are Deep Q-Net (DQN), Advantage Actor-Critic (A2C) and Simulated Annealing (SA). SA searches for globally optimal placements by iteratively refining layouts and occasionally accepting worse configurations to escape local minima. DQN, while simpler, performs robustly in environments with clear, immediate rewards. A2C handles more complex tasks but requires careful tuning and strategy design. A comparative study of A2C and DQN is found in \cite{de2024comparative}. To assess the performance of different ML methods, we use the same setup---number of places (\#P), nodes (\#N), and actions (\#A)---for each PCB. 
We have two goals: first, to find out which ML is most suitable using the proposed component centric layout, and second, to evaluate whether incorporating net information further improves placement. We perform TEWL to evaluate the performance of each method, as summarized in Table 2.
Our experiment are grouped into two categories:

\begin{itemize}
\item \textbf{Passive approach (Fig. 1a-1c)}: SA, DQN, A2C
\item \textbf{Passive + net approach (Fig. 1d)}: DQNnet
\item \textbf{Groundtruth (Fig. 1e)}: Human designed placement
\end{itemize}

\subsection{Passive Approach}

From the TEWL comparison in Table 1, A2C is in general the better performing ML than DQN or SA. However, we found DQN to be more robust than A2C for U20, U26, despite our best attempts at finetuning A2C. Apart from the most complex cases (U20 and U26), all the MLs are able to surpass GT in terms of TEWL measure. However, TEWL does not takes into account placements that have passives overlapping and routing conflicts. In Fig 1-3, by visual inspection on the MLs, we arrive at the findings in Table 3. From both TEWL and subjective visual measures, A2C is the best performing ML despite having more routing conflicts and poorer TEWL in complex cases than DQN.

\subsection{Passive + Net Approach}

By incorporating net information, DQNnet in Fig 4 see significant improvements across all the PCBs in terms of TEWL, as compared to DQN. Also, by visual inspection we saw a significant drop in overlapping passives for DQNnet vs DQN, although the routing conflicts become slightly worse. 
 
\section{Conclusions} 
In this work, we presented a novel RL approach for PCB placement with component centric layout. By encoding both passive and net ID into a unified state representation, our method enables the RL to make more informed placement decisions with respect to net proximity and passive overlap. The environment simulates PCB constraints such as fixed-grid actions and power net distribution around a main component, and the placement layout is learned through SA, DQN or A2C with a total reward function that balances Euclidean wirelength and non-overlap criteria.

\bibliographystyle{IEEEtran}
\bibliography{allmyref}

\end{document}